\documentclass[
 reprint,
superscriptaddress,
 amsmath,amssymb,
 aps,
 prx,
]{revtex4-1}

\usepackage{changes}
\usepackage{soul}

\usepackage[utf8]{inputenc}
\usepackage[english]{babel}
\usepackage{amsmath,amsfonts,amssymb}
\usepackage[T1]{fontenc}
\usepackage{graphicx}
\usepackage[colorlinks=true,bookmarks=false,citecolor=blue,urlcolor=blue]{hyperref}

\usepackage{upgreek}

\usepackage{bm}

\usepackage{enumitem}
\setlist{nolistsep}

\usepackage{xcolor}

\usepackage{letltxmacro}
\makeatletter
\let\oldr@@t\r@@t
\def\r@@t#1#2{%
	\setbox0=\hbox{$\oldr@@t#1{#2\,}$}\dimen0=\ht0
	\advance\dimen0-0.2\ht0
	\setbox2=\hbox{\vrule height\ht0 depth -\dimen0}%
	{\box0\lower0.4pt\box2}}
\LetLtxMacro{\oldsqrt}{\sqrt}
\renewcommand*{\sqrt}[2][\ ]{\oldsqrt[#1]{#2}}
\makeatother

\begin{document}
\title{Large evanescently-induced Brillouin scattering at the surrounding of a nanofibre}

\affiliation{Group for Fibre Optics, Ecole Polytechnique Fédérale de Lausanne (EPFL), Lausanne, Switzerland}
\affiliation{FEMTO-ST Institute, UMR 6174, Université Bourgogne Franche-Comté, 25030 Besançon, France}
\affiliation{State Key Laboratory of High Field Laser Physics, Shanghai Institute of Optics and Fine Mechanics, CAS, Shanghai 201800, China}
\affiliation{Present address: European Molecular Biology Laboratory, Heidelberg, Germany}
\affiliation{Present address: Max Planck Institute of Quantum Optics, Garching, Germany}

\author{Fan Yang}
\email{fanyang808@gmail.com}
\affiliation{Group for Fibre Optics, Ecole Polytechnique Fédérale de Lausanne (EPFL), Lausanne, Switzerland}
\affiliation{Present address: European Molecular Biology Laboratory, Heidelberg, Germany}

\author{Flavien Gyger}
\affiliation{Group for Fibre Optics, Ecole Polytechnique Fédérale de Lausanne (EPFL), Lausanne, Switzerland}
\affiliation{Present address: Max Planck Institute of Quantum Optics, Garching, Germany}

\author{Adrien Godet}
\affiliation{FEMTO-ST Institute, UMR 6174, Université Bourgogne Franche-Comté, 25030 Besançon, France}

\author{Jacques Chrétien}
\affiliation{FEMTO-ST Institute, UMR 6174, Université Bourgogne Franche-Comté, 25030 Besançon, France}

\author{Li Zhang}
\affiliation{Group for Fibre Optics, Ecole Polytechnique Fédérale de Lausanne (EPFL), Lausanne, Switzerland}

\author{Meng Pang}
\affiliation{State Key Laboratory of High Field Laser Physics, Shanghai Institute of Optics and Fine Mechanics, CAS, Shanghai 201800, China}

\author{Jean-Charles~Beugnot}
\email{jean-charles.beugnot@femto-st.fr}
\affiliation{FEMTO-ST Institute, UMR 6174, Université Bourgogne Franche-Comté, 25030 Besançon, France}

\author{Luc Thévenaz}
\affiliation{Group for Fibre Optics, Ecole Polytechnique Fédérale de Lausanne (EPFL), Lausanne, Switzerland}

\date{\today}

\begin{abstract}
Brillouin scattering has been widely exploited for advanced photonics functionalities such as microwave photonics, signal processing, sensing, lasing, and more recently in micro- and nano-photonic waveguides. So far, all the works have focused on the opto-acoustic interaction driven from the core region of micro- and nano-waveguides. Here we observe, for the first time, an efficient Brillouin scattering generated by an evanescent field nearby a sub-wavelength waveguide embedded in a pressurised gas cell, with a maximum gain coefficient of $18.90 \pm 0.17$ m$^{-1}$W$^{-1}$. This gain is 11 times larger than the highest Brillouin gain obtained in a hollow-core fibre and 79 times larger than in a standard single-mode fibre. The realisation of strong free-space Brillouin scattering from a waveguide benefits from the flexibility of confined light while providing a direct access to the opto-acoustic interaction, as required in free-space optoacoustics such as Brillouin spectroscopy and microscopy. Therefore, our work creates an important bridge between Brillouin scattering in waveguides, Brillouin spectroscopy and microscopy, and opens new avenues in light-sound interactions, optomechanics, sensing, lasing and imaging. 
\end{abstract}

\maketitle

Brillouin scattering involves light-sound interactions and has been used in nonlinear optics \cite{chiao_stimulated_1964,eggleton_brillouin_2019,safavi-naeini_controlling_2019,wiederhecker_brillouin_2019}, microwave photonics \cite{marpaung_integrated_2019}, slow and fast light \cite{thevenaz_slow_2008}, lasing \cite{loh_operation_2020}, sensing \cite{yang_intense_2020} and imaging \cite {prevedel_brillouin_2019,palombo_brillouin_2019}. It has been observed in various platforms, including optical fibres \cite{ippen_stimulated_1972,dainese_stimulated_2006,beugnot_brillouin_2014,florez_brillouin_2016,yang_intense_2020}, whispering-gallery-mode resonators \cite {grudinin_brillouin_2009,tomes_photonic_2009,lee_chemically_2012,kim_non-reciprocal_2015} and integrated waveguides \cite{pant_-chip_2011,shin_tailorable_2013,van_laer_interaction_2015,yang_bridging_2018,otterstrom_silicon_2018,gundavarapu_sub-hertz_2019,gyger_observation_2020}. 

The advent of micro- and nano-photonic waveguides has recently driven a renewed interest for Brillouin scattering as a tool to perform nonlinear optics and optical signal processing in waveguides. In 2006, a photonic crystal fibre was used to greatly intensify the Brillouin interactions by confining both acoustic and optical fields to a 1 $\rm \upmu m$ microstructured core \cite{dainese_stimulated_2006}. Another interesting work proposed that Brillouin interactions could be drastically enhanced by radiation pressure on subwavelength-scale waveguide boundaries \cite{rakich_giant_2012}, and triggered intense research interest into Brillouin scattering in micro- and nano-photonic waveguides. Since the gain of the interaction is proportional to the acousto-optic overlap integral as well as to the pump power, two different approaches have been addressed to increase the Brillouin interactions: the first one focused on the design of waveguides with good optical and acoustic fields overlap \cite{dainese_stimulated_2006,pant_-chip_2011,shin_tailorable_2013,beugnot_brillouin_2014,van_laer_interaction_2015}, while the second approach used ultra-low-loss waveguides with high-Q resonant enhancement \cite{grudinin_brillouin_2009,tomes_photonic_2009,lee_chemically_2012,kim_non-reciprocal_2015,yang_bridging_2018,gundavarapu_sub-hertz_2019}. Generally, all prior works were devoted to generating efficient Brillouin scattering in the core region of the waveguide. So far, no strong Brillouin scattering has been generated from an optical waveguide into the surrounding medium located in the waveguide's vicinity.

Free-space Brillouin scattering has found applications in microscopy: in the past decade, Brillouin microscopy has been used for analysing the mechanical properties of biological samples \cite{scarcelli_confocal_2008} and hydrogel materials \cite{bailey_viscoelastic_2020}. It provides label-free, non-contact, 3D imaging capabilities at typical optical resolution. So far, Brillouin microscopy has exclusively used free-space illumination, which requires careful optical alignment and has limited throughput \cite{coucheron_high-throughput_2019}. Fluorescence microscopy based on total internal reflection in photonic waveguides \cite{diekmann_chip-based_2017} and structured illumination microscopy \cite{helle_structured_2020} solve these issues for the case of fluorescence microscopy; however, there is currently no alternative for Brillouin microscopy.

Here, we use a nanofibre waveguide and report the first observation of strong  Brillouin scattering generated by the evanescent field of a guided lightwave in pressurised carbon dioxide gas located in the immediate vicinity of the nanowaveguide. We obtain a 11-times higher peak Brillouin gain coefficient in the nanofibre gas cell compared to the highest Brillouin gain realised in a hollow-core fibre gas cell \cite{yang_intense_2020} and 79-times higher compared to that in a standard single-mode fibre. The higher gain results from a tighter optical mode confinement as well as a large overlap between the evanescent optical field and the acoustic mode in the surrounding medium. Our results pave the way for many potential applications in sensing, lasing, signal processing, etc and suggest a way to realise a novel waveguide-based Brillouin spectroscopy and microscopy that no longer requires free-space illumination.

\vspace{5mm}

\begin{figure*}[t!]
	\includegraphics[width = 18cm]{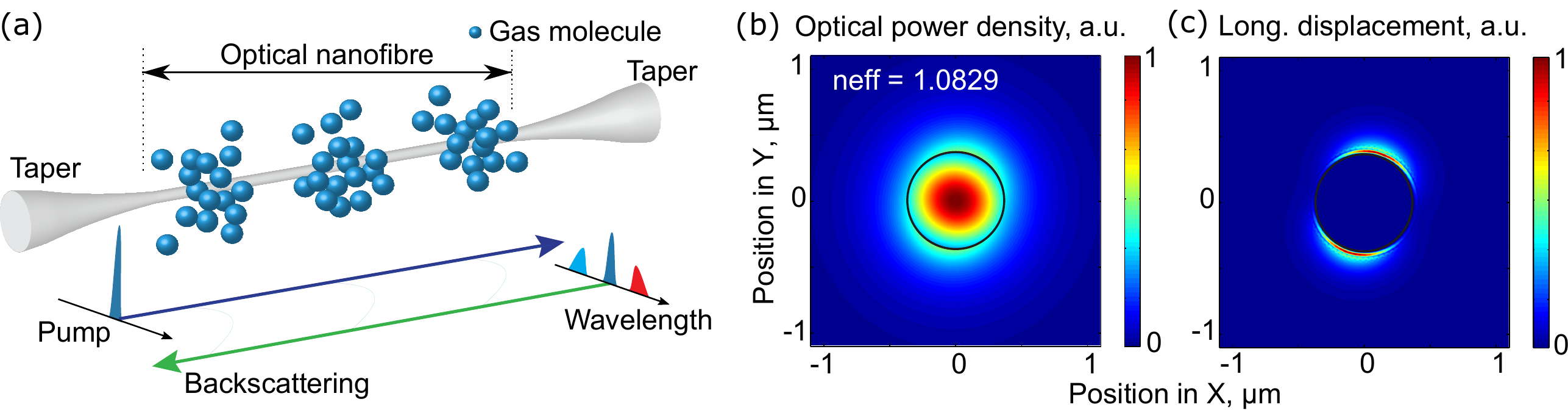}
	\caption{\textbf{Brillouin scattering in a nanofibre gas cell.} \textbf{(a)} Conceptual view of the Brillouin scattering in a nanofibre gas cell. Pump (in green) goes from left to right and the backscattered light includes Rayleigh scattering from the same frequency and frequency down-shifted Stokes (in red) and frequency up-shifted anti-Stokes (in blue) Brillouin scattering. \textbf{(b)} Computed spatial optical power distribution of the fundamental optical mode at a wavelength of 1550 nm for a 740 nm diameter nanofibre surrounded with 40 bar CO$_2$ gas. The effective mode refractive index is 1.0829. \textbf{(c)} Computed longitudinal displacement of the elastic wave at 350 MHz as a result of backward Brillouin scattering in the surrounding gas. The black circles in (b) and (c) show the silica nanofibre boundary. See Methods for the simulation details.}
	\label{fig_Principle}
\end{figure*}

\noindent {\large\textbf{Results}}

\noindent \textbf{Operation scheme.} 
Brillouin scattering is an inelastic process within a medium involving a net energy transfer from the optical fields to the medium or vice versa. Spontaneous Brillouin scattering occurs when light scatters from thermally-excited sound waves, giving rise to frequency-shifted Stokes (i.e. phonon generation) and anti-Stokes waves (phonon annihilation). A conceptual illustration of Brillouin scattering around a nanowaveguide is shown in Fig. \ref{fig_Principle}(a). In a waveguide, only photons scattered in the backward and forward directions are guided and hence detected. Here we focus on backward Brillouin scattering. The backward scattering efficiency is maximised when phase matching between all interacting waves is satisfied. For the Stokes process, an incident pump photon of frequency $\omega_{\rm P}$ is converted to a lower frequency Stokes photon of frequency $\omega_{\rm S}$ through the scattering from the acoustic wave and thereby creating a phonon of angular frequency $\Omega$. The phase matching condition requires $\omega_{\rm P} = \omega_{\rm S} + \Omega$ and $k_{\rm P} = k_{\rm S} + q$, where $k_{\rm P}$, $k_{\rm S}$ and $q$ are the wave vectors of the pump, Stokes, and phonon modes, respectively. In this case, the Brillouin frequency shift $\nu_{\rm B}$, which is the frequency difference between pump and Stokes waves under phase matching condition, is given by \cite{boyd_nonlinear_2008}:
\begin{equation}
\label{eq:phasematching}
\nu_{\rm B}=2n_{\rm eff}v_{\rm a}/\lambda,
\end{equation}

where $n_{\rm eff}$ is the effective refractive index of the optical mode, $v_{\rm a}$ is the acoustic velocity of the medium - the surrounding gas in our case - and $\lambda$ is the pump wavelength in vacuum. 

The small nanofibre dimensions (740 nm in this work) compared to the optical wavelength results in 58\% of the light field intensity propagating outside the nanofibre, which can interact with the surrounding gas or other fluid material. This evanescent field is visible in the finite-element simulation of the power distribution of the fundamental optical mode illustrated in Fig. \ref{fig_Principle}(b). The gas molecular displacement of the acoustic mode along the fibre axis direction is shown in Fig. \ref{fig_Principle}(c) (see Methods for the simulation details). 

\vspace{3mm}

\begin{figure}[hbt!]
		\includegraphics[width = 8.5cm]{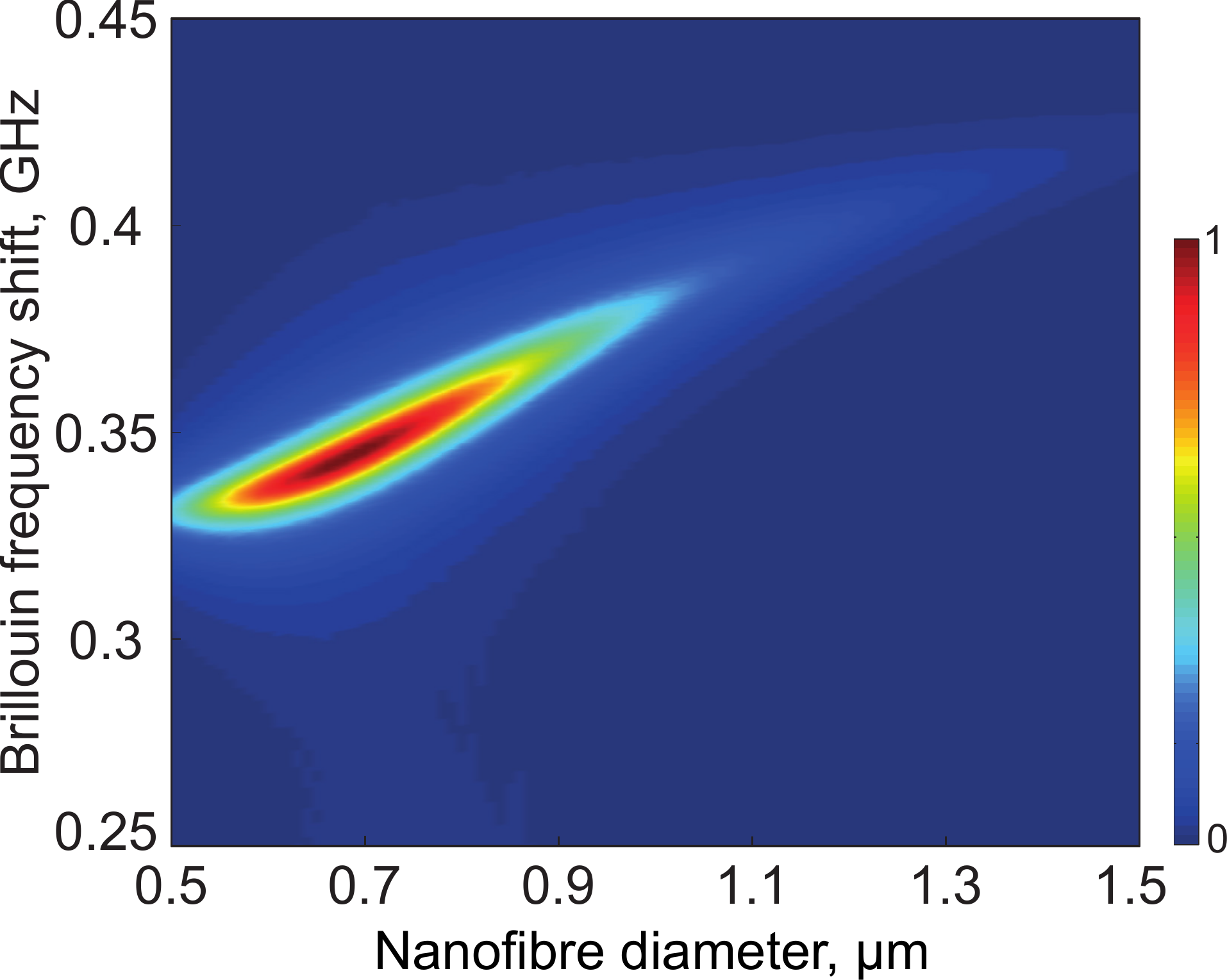}
		\caption{3D calculation of Brillouin spectra in silica nanofibres surrounded by 40 bar CO$_2$ for a diameter ranging from 500 nm to 1.5 $\rm \upmu$m.}
		\label{spectre_lin_3D}
\end{figure}

\begin{figure*}[ht!]
		\includegraphics[width = 1\linewidth]{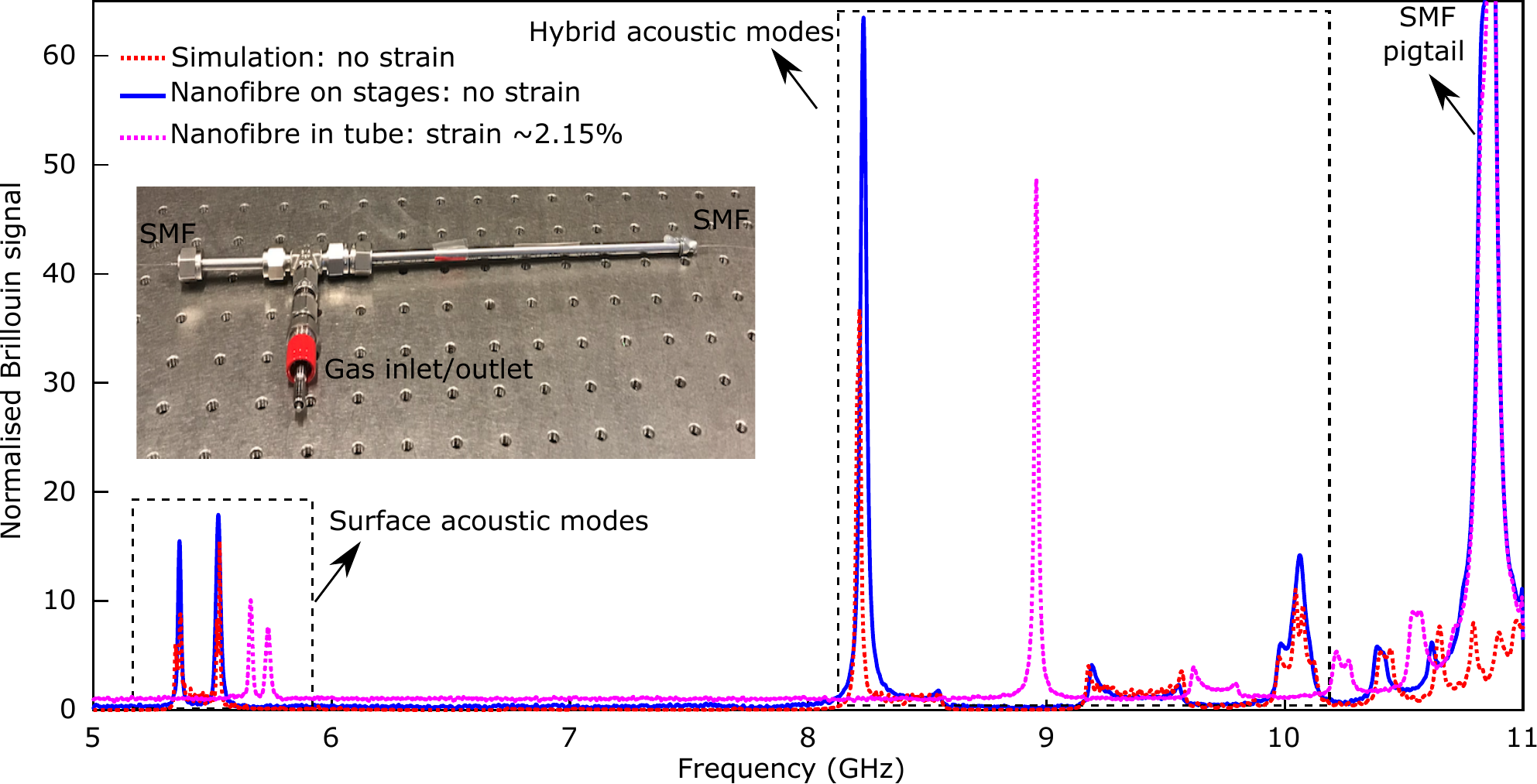}
		\caption{\textbf{Backward Brillouin scattering in the silica material of the nanofibre during fabrication process.} The backward Brillouin scattering of the nanofibre is measured with a set-up we developed in 2017 \cite{godet_brillouin_2017}. The Brillouin signal is normalised to the background noise of the electrical spectrum analyser when there is no pump. The Brillouin spectra from the surface acoustic modes and the hybrid acoustic modes of the silica nanofibre during fabrication are calculated by the model in \cite{godet_brillouin_2017}. The blue line shows the measured Brillouin spectrum of the nanofibre on the stages after tapering. The dotted red line shows the simulated Brillouin spectrum of the nanofibre with a waist diameter of 740 nm. The dotted purple line shows the measured Brillouin spectrum of the nanofibre after packaging the nanofibre in a metallic tube with a 2.15\% strain. Note that the signal from SMF pigtail is not taken into account in the simulation and any response from interactions in the gas is not visible in the frequency range covered by the set-up. The Brillouin scattering from the standard single-mode fibre pigtail of the nanofibre is also observed at a frequency of 10.86 GHz. The inset shows the fabricated nanofibre gas cell with the untapered SMF segments exiting the gas cell at each end. Gases can be pressurised in or vacuum pumped out through the gas inlet/outlet port.}
		\label{fig_Brillouin_nanofibre}
\end{figure*}

\noindent \textbf{Nanofibre gas cell.}
The Brillouin backscattering spectrum from an elastic wave generated in the gas surrounding the silica nanofibre can be neatly observed. The large evanescent field creates a strong electrostrictive force spatially distributed around the nanofibre. The elastic wave in the gas generated via electrostriction is confined by the evanescent optical field.
Figure \ref{spectre_lin_3D} shows a numerical calculation giving the map of the Brillouin spectrum in 40 bar CO$_2$ surrounding a nanofibre with diameters ranging from 0.5 to 1.5 $\upmu$m. This calculation indicates the geometrical sizes of the nanofibre maximising the light-sound interaction in the surrounding gas, which is a trade-off between the optical mode's effective overlap and the optical intensity in the evanescent field. The optimal diameter of the nanofibre maximising Brillouin scattering by the evanescent field is calculated to be 740 nm. 

The detailed fabrication process of the nanofibre gas cell is described in Methods. We infer the waist diameter of the nanofibre with a few nanometre sensitivity by mapping the backscattered Brillouin spectrum along the optical fibre taper and fitting with numerical simulations of the elastodynamic equations (a technique we developed in 2017 \cite{godet_brillouin_2017}). The fabricated nanofibre sample presented in this work shows a waist diameter of 740 nm $\pm$ 5 nm, a waist length of 10 cm, two adiabatic transitions of 79.4 mm $\pm$ 5 mm of total length and an insertion loss of 0.14 dB. The backscattered Brillouin signal of the nanofibre is shown in Fig. \ref{fig_Brillouin_nanofibre}. The blue and dotted red lines in Fig. \ref{fig_Brillouin_nanofibre} are respectively the experimental and numerical simulation results in the nanofibre after the tapering process. The experimental Brillouin spectrum, including the surface acoustic modes and hybrid acoustic modes, matches perfectly with the numerical simulation for a nanofibre with a diameter of 740 nm $\pm$ 5 nm. After fabrication, the nanofibre is packaged in a simple metallic tube as a gas cell while applying a controlled strain. The measured Brillouin spectrum after applying the strain is shown in the dotted purple line of Fig. \ref{fig_Brillouin_nanofibre}. As a result of the strain, the resonances due to surface acoustic modes and hybrid acoustic modes both shift to higher frequencies, perfectly matching with our previous analysis \cite{godet_nonlinear_2019}.

\begin{figure*}[t!]
	\includegraphics[width = 0.8\linewidth]{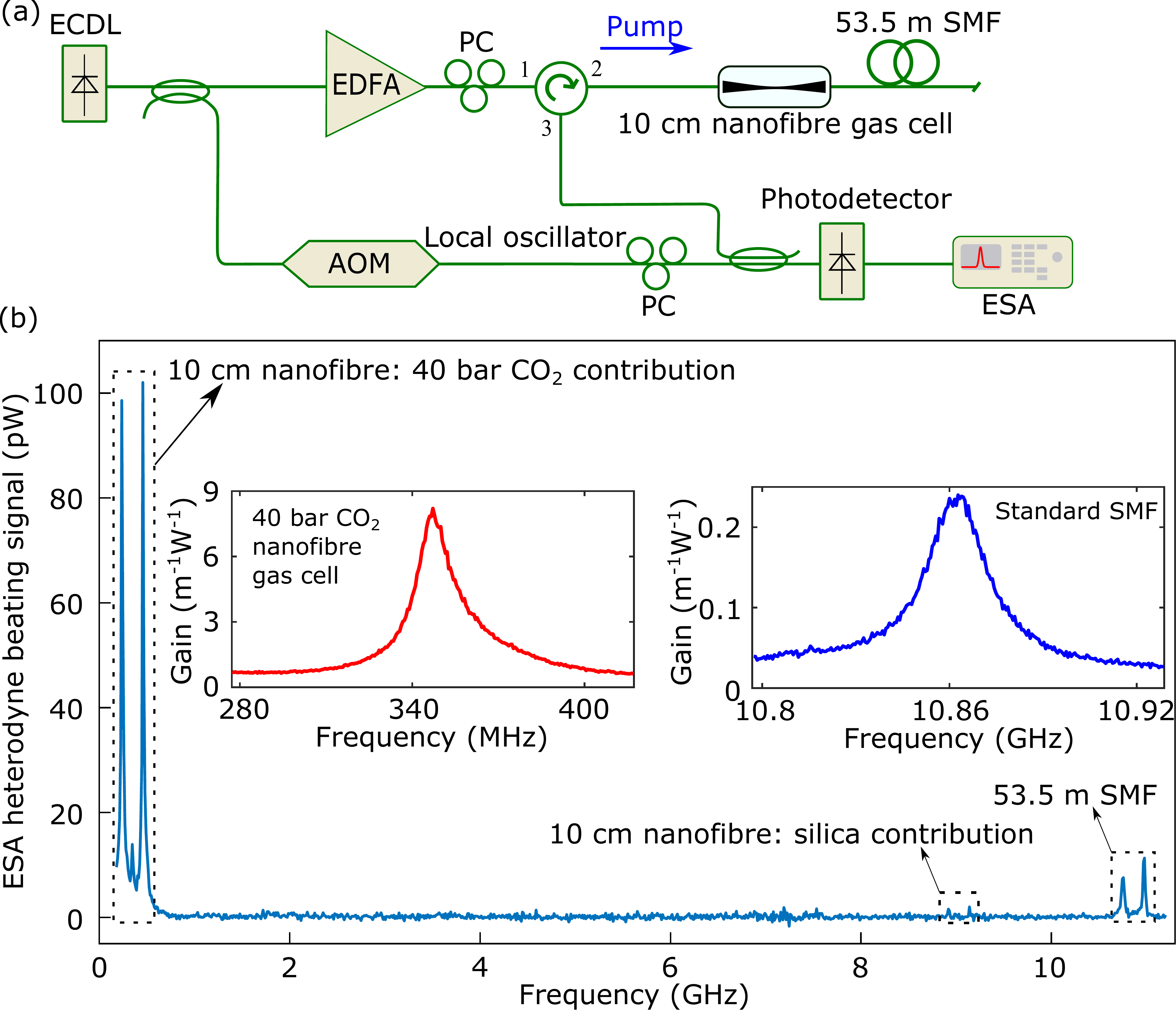}
	\caption{\textbf{Backward Brillouin scattering in the nanofibre gas cell.} \textbf{(a)} Heterodyne experimental set-up. To make a comparison between peak Brillouin gains, a 53.5 m single-mode fibre (SMF) is connected to the far end of the nanofibre. The output of a continuous-wave external-cavity diode laser (ECDL) is split: one branch is amplified by an erbium-doped fibre amplifier (EDFA) and used to pump the nanofibre as well as the SMF. The other branch is up-shifted in frequency (+110 MHz) by an acousto-optic modulator (AOM) and combined with the backward spontaneous Brillouin scattering for heterodyne mixing. The total length of the nanofibre (including the adiabatic tapering transition and the nanofibre waist) is 179.4 mm $\pm$ 5 mm with a 10 cm long 740 nm $\pm$ 5 nm diameter nanofibre. Note that the total insertion loss of the nanofibre is 1 dB including the 0.14 dB tapering-induced loss, the splicing loss of the nanofibre to two fibre pigtails at each end, as well as the loss of the two angled-physical contact connectors. The heterodyne spectrum is measured by a photodetector combined with a radio-frequency electrical spectrum analyser (ESA). The AOM is used to separate the Stokes and anti-Stokes Brillouin scatterings. The polarisations of the pump light and the local oscillator are adjusted using two polarisation controllers (PCs) to achieve the highest heterodyne response. \textbf{(b)} Heterodyne beating spectra of the 10 cm nanofibre gas cell and the appended 53.5 m SMF. The Brillouin frequency shifts of the 40 bar CO$_2$ in the 10 cm nanofibre gas cell, the silica contribution in the 10 cm nanofibre gas cell as well as the 53.5 m SMF are 0.34 GHz, 9 GHz and 10.86 GHz respectively. The red and blue lines in the inset of \textbf{(b)} are the backward Stokes Brillouin gain spectra for the 40 bar CO$_2$ nanofibre gas cell and for the SMF, respectively. The horizontal scales span over the same frequency width for a good comparison.}
	\label{fig_NF_SMF_spectrum}
\end{figure*}

\begin{figure*}[hbt!]
		\includegraphics[width = 1\linewidth]{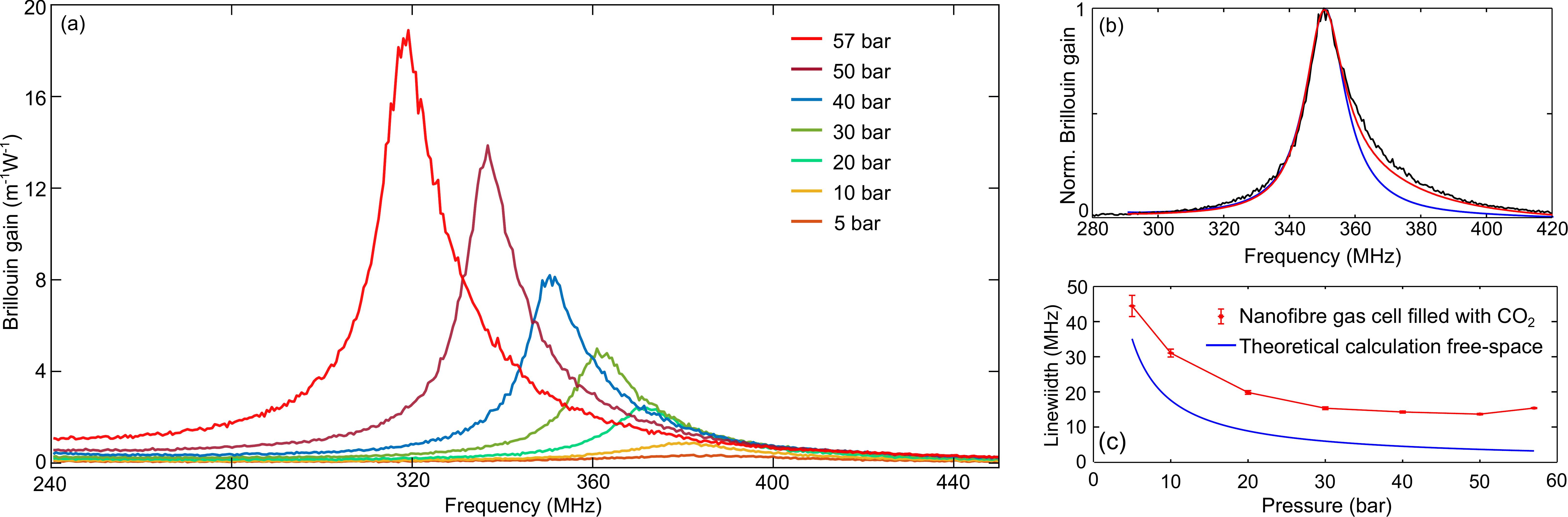}
		\caption{\textbf{Experimental Brillouin gain spectra in the nanofibre placed in a gas cell filled with CO$_2$.} 
		\textbf{(a)} Measured Brillouin gain spectra along the nanofibre gas cell filled with CO$_2$ at different pressures. 
		\textbf{(b)} Normalised Brillouin gain spectra in 40 bar CO$_2$ obtained in a nanofibre with a waist of 740 nm. The blue line shows the theoretical Brillouin spectrum at the nanofibre waist; the red line shows the compound theoretical Brillouin spectrum along the nanofibre waist and the tapered region. The black line shows the experimental Brillouin spectrum along the nanofibre in the gas cell filled with 40 bar CO$_2$.
		\textbf{(c)} Measured Brillouin linewidth defined as twice the value of the left half-width measured spectrum using a Lorentzian fitting on the experimental curves in (a). Each red data point shows the estimated mean and the RMS error bar for a total of 10 measurements at each pressure. The blue line shows the theoretical Brillouin linewidth of free-space CO$_2$ gas at different pressures.}
		\label{fig_CO2_pressures}
\end{figure*}

\vspace{3mm}
\noindent \textbf{Evanescently-induced Brillouin scattering in the nanofibre gas cell.} The Brillouin gain coefficient and spectrum can be obtained from a spontaneous Brillouin scattering measurement \cite{boyd_noise_1990}. The detailed calculation and calibration of the Brillouin gain coefficient from the spontaneous Brillouin measurements are described in Supplementary Section 1. All the experiments were performed at an environmental temperature of 24 $\pm~1^{\circ}$C. Figure \ref{fig_NF_SMF_spectrum}(a) shows the detailed experimental implementation. The pump light is amplified by an erbium-doped fibre amplifier and launched through a circulator into the nanofibre gas cell. A 53.5~m single-mode fibre (SMF) is appended to the nanofibre and provides a direct comparative response for the Brillouin gain. The backward spontaneous Brillouin scattering signal is measured using heterodyne beating with a frequency-upshifted fraction of the pump laser (as a local oscillator) and detected by a radio-frequency electrical spectrum analyser via a photodetector. The beating spectra from the 40~bar CO$_2$ contribution in the 10~cm nanofibre gas cell, from the silica contribution in the 10~cm nanofibre as well as from the 53.5 m SMF are shown in Fig.~\ref{fig_NF_SMF_spectrum}(b). The Brillouin frequency shift for the 40 bar CO$_2$ contribution, silica contribution and SMF contribution are 340~MHz, 9~GHz and 10.86~GHz respectively. Note that the 9~GHz signal is from the silica nanofibre hybrid acoustic mode which has been analysed in Fig.~\ref{fig_Brillouin_nanofibre}. The +110~MHz frequency-shifted local oscillator splits the Stokes and anti-Stokes components. The right peaks identify the Stokes scattering while the left peaks originate from the anti-Stokes scattering. Remarkably, we can observe that the peak Brillouin signal for 10 cm~nanofibre filled with 40~bar CO$_2$ is $\sim$ 10 times higher than that of the 53.5~m SMF.

The red and blue lines in the inset of Fig. \ref{fig_NF_SMF_spectrum} (b) show the Brillouin gain spectra of the 40 bar CO$_2$ nanofibre gas cell and the SMF respectively, when the polarisations are tuned to maximise the specific Brillouin signal. The peak Brillouin gain coefficient for the nanofibre gas cell filled with 40 bar CO$_2$ is $8.20 \pm 0.08$ m$^{-1}$W$^{-1}$ which is 34 times higher than that of the SMF (0.24 m$^{-1}$W$^{-1}$). The error bar is calculated from the standard deviation of the peak gain over 10 measurements. Note that the peak Brillouin gain coefficient for the SMF obtained from our spontaneous scattering measurement is in good agreement with the standard value for this type of fibre \cite{motil_invited_2016}, which confirms the solidity of our test bench and calibrations based on spontaneous scattering. The shape of the Brillouin gain spectrum of the standard SMF is a fully symmetric Lorentzian distribution while the shape of the Brillouin gain spectrum of the 40 bar CO$_2$ nanofibre gas cell is evidently asymmetric, as explained in the next section.

\begin{figure*}[hbt!]
		\includegraphics[width = 14cm]{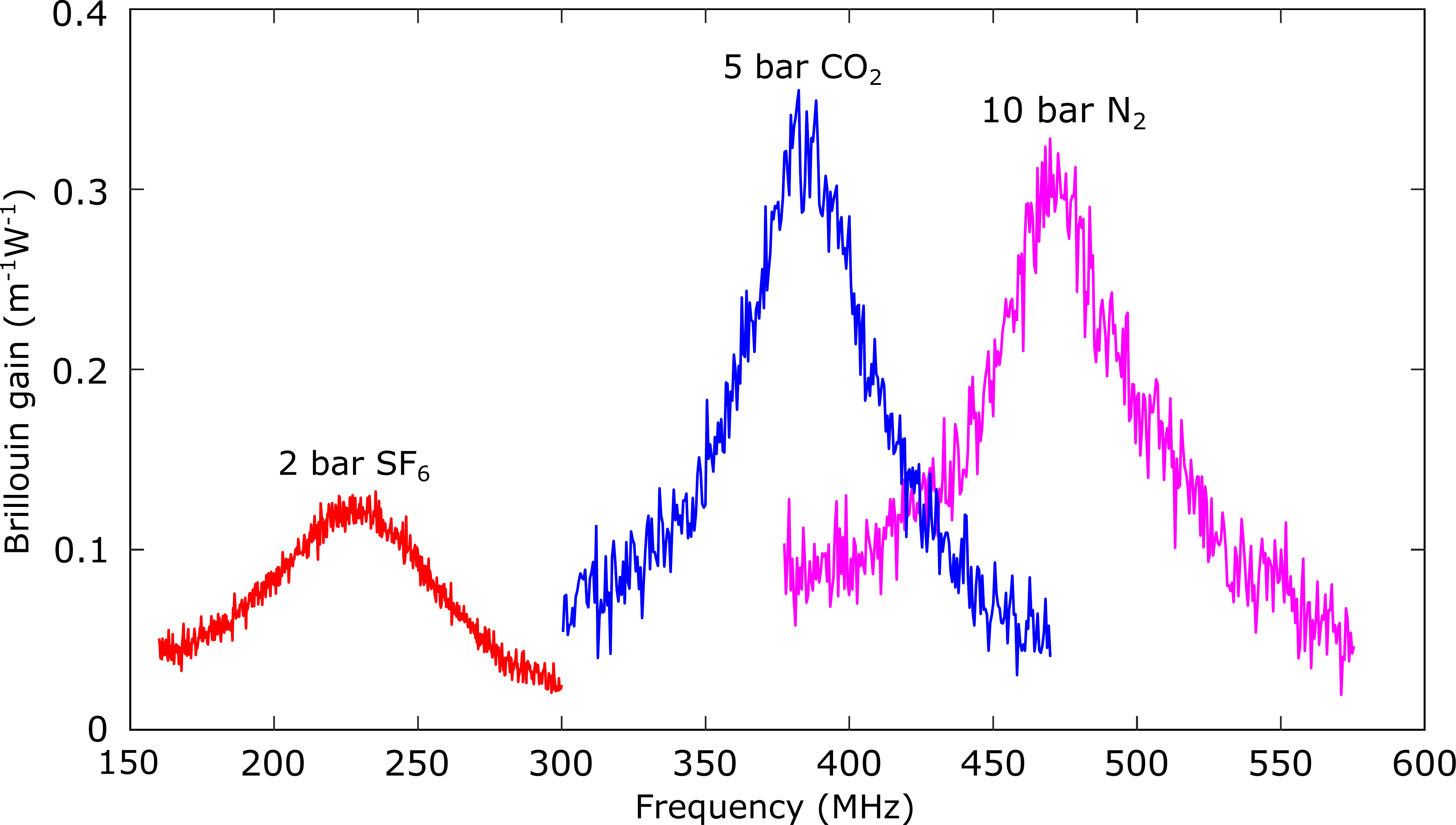}
		\caption{\textbf{Experimental Brillouin gain spectra from the nanofibre gas cell filled with different types of gas.} Measured Brillouin gain spectra for 2 bar SF$_6$, 5 bar CO$_2$ and 10 bar N$_2$. For the CO$_2$ and N$_2$ measurements, the noise equivalent bandwidth of the electrical spectrum analyser (ESA) is 1 kHz, while for the SF$_6$ measurement, the noise equivalent bandwidth of the ESA is 100 Hz.}
		
		\label{fig_different_gases}
\end{figure*}

\vspace{3mm}
\noindent \textbf{Different gas pressures.}
Figure \ref{fig_CO2_pressures}(a) shows the Brillouin gain coefficient as a function of the CO$_2$ pressure in the nanofibre gas cell. The peak gain coefficient increases with the pressure while the linewidth is simultaneously reduced, in agreement with the model presented in \cite{yang_intense_2020}. The peak gain coefficient is $18.90 \pm 0.17$ m$^{-1}$W$^{-1}$ at 57 bar which is 79 times stronger than that in a SMF. The error bar is calculated from the standard deviation of the peak gain for 10 measurements.

It is interesting to see in Fig. \ref{fig_CO2_pressures}(a) that the shape of the Brillouin spectra is skewed. This asymmetry results from the contribution of the tapered transition region, since this segment shows a position-varying evanescent optical field that keeps interacting with the sound wave in the gas. Let us take 40 bar CO$_2$ as an example and analyse its spectrum in Fig. \ref{fig_CO2_pressures}(b). The blue line is the theoretical Brillouin spectrum in 40 bar CO$_2$ (along the homogeneous nanofibre section) and the red line is the theoretical Brillouin spectrum including the uniform nanofibre section as well as the tapered region (see Supplementary Section 2 for the detailed analysis). The tapered region only contributes to the higher frequency region because its gradually larger diameter increases the effective refractive index, leading to a larger Brillouin frequency shift, as indicated by the phase-matching condition given in Eq. \ref{eq:phasematching}. So the tapered region presents a higher Brillouin frequency shift and makes the Brillouin spectrum of the nanofibre slightly skewed. The theoretical Brillouin spectrum taking into account this asymmetry matches well with our experimental results. 

Figure \ref{fig_CO2_pressures}(c) shows the measured Brillouin linewidth for the nanofibre gas cell filled with CO$_2$ at different pressures as well as the theoretical estimation of the free-space gas Brillouin linewidth. Here, to exclude the contribution from the tapered regions, the Brillouin linewidth for the nanofibre gas cell is defined as twice the value of the left half-width-half-maximum of the spectrum fitted with a Lorentzian (i.e. considering only the left half of the spectrum). The measured Brillouin linewidth decreases with pressure in the range from 5 to 50 bar. It then increases from 50 to 57 bar, as a result of the close vicinity of the gas-liquid phase transition at room temperature, since the acoustic damping increases when the gas phase is approaching the liquid transition. It should be mentioned that the measured Brillouin linewidth for the nanofibre at a specific pressure is $\sim 10$ MHz larger than the free space gas Brillouin linewidth. This linewidth broadening is thought to originate from the coupling of the light field with a continuum of free-space modes \cite{poulton2013acoustic}, unlike in our simulation in which we only considered one acoustic mode propagating longitudinally.

\vspace{3mm}
\noindent \textbf{Different types of gas.} We then carried out the study of Brillouin scattering in a nanofibre gas cell filled with different types of gas, namely CO$_2$, sulfur hexafluoride (SF$_6$) and nitrogen (N$_2$). The Brillouin gain spectra for these three gases at specific pressures are shown in Fig. \ref{fig_different_gases}. This result shows the possibilities of the evanescent Brillouin scattering of our nanofibre to be used for Brillouin spectroscopy and gas analysis as well as the flexibility of our platform in tailoring the Brillouin gain at will.

\vspace{5mm}
\noindent {\large\textbf{Discussion}} 

\noindent In this work, we have demonstrated the generation of a strong Brillouin scattering driven by the evanescent field of a nanofibre placed in a gas-filled cell. The nanofibre is obtained by fused-tapering of a standard single mode fibre, so that the mode size transition to a sub-micron dimension can be adiabatically realised with very low added loss: this is a crucial advantage for the implementation of such a gain stage in photonic systems. The tighter light confinement in nanofibre compared with hollow-core fibre results in a measured Brillouin gain coefficient in our 40 bar CO$_2$ gas filled nanofibre gas cell 5-times larger than in a hollow-core fibre under the same pressure \cite{yang_intense_2020}, but it is expected that higher gains can be achieved by using noble gases (e.g. xenon) \cite{eden_optical_1974}. Using our nanofibre, we can increase the pressure close to the gas-liquid phase transition of CO$_2$ (64 bar at ambient temperature) with negligible absorption loss because of the relatively shorter length of the nanofibre and smaller percentage of light interact with the gas molecules, in contrast with hollow core fibres where light molecular absorption turns out to limit the maximum pressure. In these conditions, we have achieved a peak Brillouin gain coefficient of $18.90 \pm 0.17$ m$^{-1}$W$^{-1}$ at 57 bar, which is 11 times higher than that obtained in a hollow-core fibre and is 79 times higher than that in a SMF. Furthermore, unlike in hollow-core fibres \cite{yang_intense_2020} where both optical and acoustic modes are confined in the hollow core, this work has demonstrated the first possibility to probe the acoustic wave in the surrounding medium of a waveguide using an all-optical method thanks to the strong Brillouin scattering generated by the evanescent field of the waveguide.

This feature offers the attractive possibility to realise compact optical amplification stages, since a gain similar to that generated over several tens of meter of standard fibre can be obtained over a few centimetres along the nanofibre, with a very minor insertion loss. The concept may be extended to any planar waveguiding structure showing a substantial evanescent field and can be the essence of gain blocks in photonic integrated circuits. For instance, a suspended \cite {van_laer_interaction_2015} or slot nano-waveguide \cite {almeida_guiding_2004} could be used for strong light-sound interactions in gas. Various functional optical devices and sensors could be generated based on this platform.

Our nanofibre gas cell can also be used for novel approaches in pressure and temperature sensing. For instance, the measured Brillouin frequency shift along the nanofibre gas cell filled with CO$_2$ at different pressures (shown in Fig. S2 in Supplementary Section 3) shows a pressure sensitivity of $\sim$ -1 MHz/bar. This large sensitivity enables our nanofibre gas cell to be used for pressure measurement over a range from 5 to 57 bar. Temperature variations also change the gas acoustic velocity  \cite{yang_intense_2020} and hence the Brillouin frequency shift in the nanofibre gas cell. 

In order to estimate the capability of our nanofibre waveguide for microscopy and spectroscopy of biological samples, we have simulated the Brillouin gain spectrum of a nanofibre immersed in water. Since the acoustic velocities in the surrounding medium (e.g. $\sim 300$ m/s in gas and 1500 m/s in water) and that of the solid silica nanofibre (6000 m/s) are very different, the respective Brillouin frequency shifts in the surrounding medium and the nanofibre are well separated. The peak Brillouin gain is calculated to be 1 m$^{-1}$W$^{-1}$ for a 450 nm diameter nanofibre immersed in water, which is 8 times larger than that in 2 bar of SF$_6$ shown in Fig. \ref{fig_different_gases}. The detailed simulation is described in Fig. S3 in Supplementary Section 4. The accessibility of the acoustic wave, present in the surrounding medium and thus external to the waveguide, and the high Brillouin gain of a nanofibre immersed in water opens new perspectives and creates an important bridge between Brillouin scattering in waveguide and Brillouin microscopy and spectroscopy. So far, Brillouin microscopy and spectroscopy have exclusively relied on free-space light excitation. For Brillouin spectroscopy \cite{bailey_viscoelastic_2020}, our platform provides an efficient way for light illumination and Brillouin scattering signal collection. For Brillouin microscopy, our work suggests a novel imaging modality, namely waveguide-illumination Brillouin microscopy, which inherits many benefits of on-chip based microscopy \cite{diekmann_chip-based_2017,helle_structured_2020}, such as separation between the illumination and detection light paths, facilitated alignment and high throughput \cite{coucheron_high-throughput_2019}. 

\vspace{5mm}
\noindent \textbf{Methods}

\noindent \textbf{Simulations.}
The transverse dimension of the silica nanofibre is close to the acoustic wavelength. Therefore, in such a waveguide, the boundaries conditions induce a strong coupling between longitudinal, shear and surface elastic components. The generation of a confined elastic wave in the gas-silica rod assembling by the two components of optical field (guided and evanescent) is calculated by using the elasto-dynamic equation driven by the electrostrictive stress \cite{Laude_2018,Beugnot_2012}.
All calculations for CO$_2$ at 40 bar are realised using: an optical wavelength of 1550 nm, a refractive index of 1.01804 \cite{yang_intense_2020}, an acoustic velocity of 250 $\rm m/s$, and a density of 71.35 $\rm kg/m^3$. To exclude the contribution from the tapered regions, the Brillouin linewidth used in the simulation for 40 bar CO$_2$ is 14.2 MHz which is defined as twice the value of the left half-width-half-maximum of the spectrum fitted with a Lorentzian (i.e. considering only the left half of the spectrum).

\noindent \textbf{Fabrication of the nanofibre gas cell.}
The nanofibre is fabricated by tapering a standard single-mode fibre (SMF) using the heat-brush technique. To package the nanofibre into the gas cell, we applied the following procedure: first, we move the nanofibre sample from the tapering translation stage to a fixed-length stage with two fibre clamps on both ends of the SMF pigtails. Second, we orient the stage perpendicular to the ground, release the lower fibre clamp, insert the lower end of the nanofibre sample into a metallic tube, to finally hermetically glue the lower end of the tube. Third, to avoid any contact of the nanofibre with the tube wall, we apply a 2.15\% strain on the nanofibre. We then apply glue on the other side while precisely controlling the strain and monitoring the backward Brillouin spectrum.

\vspace{3mm}
\noindent \textbf{Acknowledgments.}
The authors would like to acknowledge the support by the Swiss National Science Foundation (SNSF) under grant No. 159897 and 178895, and the financial support of Agence Nationale de la Recherche (ANR-16-CE24-0010-03), EIPHI Graduate School (contract ANR-17-EURE-0002) and Bourgogne-Franche-Comté Region.

\noindent \textbf{Data Availability Statement}: The code and data used to produce the plots within this work will be released on the repository \texttt{Zenodo} upon publication of this preprint.

\bibliographystyle{apsrev4-1}
\bibliography{ms}

\begin{thebibliography}{41}%
\makeatletter
\providecommand \@ifxundefined [1]{%
 \@ifx{#1\undefined}
}%
\providecommand \@ifnum [1]{%
 \ifnum #1\expandafter \@firstoftwo
 \else \expandafter \@secondoftwo
 \fi
}%
\providecommand \@ifx [1]{%
 \ifx #1\expandafter \@firstoftwo
 \else \expandafter \@secondoftwo
 \fi
}%
\providecommand \natexlab [1]{#1}%
\providecommand \enquote  [1]{``#1''}%
\providecommand \bibnamefont  [1]{#1}%
\providecommand \bibfnamefont [1]{#1}%
\providecommand \citenamefont [1]{#1}%
\providecommand \href@noop [0]{\@secondoftwo}%
\providecommand \href [0]{\begingroup \@sanitize@url \@href}%
\providecommand \@href[1]{\@@startlink{#1}\@@href}%
\providecommand \@@href[1]{\endgroup#1\@@endlink}%
\providecommand \@sanitize@url [0]{\catcode `\\12\catcode `\$12\catcode
  `\&12\catcode `\#12\catcode `\^12\catcode `\_12\catcode `\%12\relax}%
\providecommand \@@startlink[1]{}%
\providecommand \@@endlink[0]{}%
\providecommand \url  [0]{\begingroup\@sanitize@url \@url }%
\providecommand \@url [1]{\endgroup\@href {#1}{\urlprefix }}%
\providecommand \urlprefix  [0]{URL }%
\providecommand \Eprint [0]{\href }%
\providecommand \doibase [0]{http://dx.doi.org/}%
\providecommand \selectlanguage [0]{\@gobble}%
\providecommand \bibinfo  [0]{\@secondoftwo}%
\providecommand \bibfield  [0]{\@secondoftwo}%
\providecommand \translation [1]{[#1]}%
\providecommand \BibitemOpen [0]{}%
\providecommand \bibitemStop [0]{}%
\providecommand \bibitemNoStop [0]{.\EOS\space}%
\providecommand \EOS [0]{\spacefactor3000\relax}%
\providecommand \BibitemShut  [1]{\csname bibitem#1\endcsname}%
\let\auto@bib@innerbib\@empty
\bibitem [{\citenamefont {Chiao}\ \emph {et~al.}(1964)\citenamefont {Chiao},
  \citenamefont {Townes},\ and\ \citenamefont
  {Stoicheff}}]{chiao_stimulated_1964}%
  \BibitemOpen
  \bibfield  {author} {\bibinfo {author} {\bibfnamefont {R.~Y.}\ \bibnamefont
  {Chiao}}, \bibinfo {author} {\bibfnamefont {C.~H.}\ \bibnamefont {Townes}}, \
  and\ \bibinfo {author} {\bibfnamefont {B.~P.}\ \bibnamefont {Stoicheff}},\
  }\href {\doibase 10.1103/PhysRevLett.12.592} {\bibfield  {journal} {\bibinfo
  {journal} {Physical Review Letters}\ }\textbf {\bibinfo {volume} {12}},\
  \bibinfo {pages} {592} (\bibinfo {year} {1964})}\BibitemShut {NoStop}%
\bibitem [{\citenamefont {Eggleton}\ \emph {et~al.}(2019)\citenamefont
  {Eggleton}, \citenamefont {Poulton}, \citenamefont {Rakich}, \citenamefont
  {Steel},\ and\ \citenamefont {Bahl}}]{eggleton_brillouin_2019}%
  \BibitemOpen
  \bibfield  {author} {\bibinfo {author} {\bibfnamefont {B.~J.}\ \bibnamefont
  {Eggleton}}, \bibinfo {author} {\bibfnamefont {C.~G.}\ \bibnamefont
  {Poulton}}, \bibinfo {author} {\bibfnamefont {P.~T.}\ \bibnamefont {Rakich}},
  \bibinfo {author} {\bibfnamefont {M.~J.}\ \bibnamefont {Steel}}, \ and\
  \bibinfo {author} {\bibfnamefont {G.}~\bibnamefont {Bahl}},\ }\href {\doibase
  10.1038/s41566-019-0498-z} {\bibfield  {journal} {\bibinfo  {journal} {Nature
  Photonics}\ }\textbf {\bibinfo {volume} {13}},\ \bibinfo {pages} {664}
  (\bibinfo {year} {2019})}\BibitemShut {NoStop}%
\bibitem [{\citenamefont {Safavi-Naeini}\ \emph {et~al.}(2019)\citenamefont
  {Safavi-Naeini}, \citenamefont {Thourhout}, \citenamefont {Baets},\ and\
  \citenamefont {Laer}}]{safavi-naeini_controlling_2019}%
  \BibitemOpen
  \bibfield  {author} {\bibinfo {author} {\bibfnamefont {A.~H.}\ \bibnamefont
  {Safavi-Naeini}}, \bibinfo {author} {\bibfnamefont {D.~V.}\ \bibnamefont
  {Thourhout}}, \bibinfo {author} {\bibfnamefont {R.}~\bibnamefont {Baets}}, \
  and\ \bibinfo {author} {\bibfnamefont {R.~V.}\ \bibnamefont {Laer}},\ }\href
  {\doibase 10.1364/OPTICA.6.000213} {\bibfield  {journal} {\bibinfo  {journal}
  {Optica}\ }\textbf {\bibinfo {volume} {6}},\ \bibinfo {pages} {213} (\bibinfo
  {year} {2019})}\BibitemShut {NoStop}%
\bibitem [{\citenamefont {Wiederhecker}\ \emph {et~al.}(2019)\citenamefont
  {Wiederhecker}, \citenamefont {Dainese},\ and\ \citenamefont
  {Mayer~Alegre}}]{wiederhecker_brillouin_2019}%
  \BibitemOpen
  \bibfield  {author} {\bibinfo {author} {\bibfnamefont {G.~S.}\ \bibnamefont
  {Wiederhecker}}, \bibinfo {author} {\bibfnamefont {P.}~\bibnamefont
  {Dainese}}, \ and\ \bibinfo {author} {\bibfnamefont {T.~P.}\ \bibnamefont
  {Mayer~Alegre}},\ }\href {https://aip.scitation.org/doi/10.1063/1.5088169}
  {\bibfield  {journal} {\bibinfo  {journal} {APL Photonics}\ }\textbf
  {\bibinfo {volume} {4}},\ \bibinfo {pages} {071101} (\bibinfo {year}
  {2019})}\BibitemShut {NoStop}%
\bibitem [{\citenamefont {Marpaung}\ \emph {et~al.}(2019)\citenamefont
  {Marpaung}, \citenamefont {Yao},\ and\ \citenamefont
  {Capmany}}]{marpaung_integrated_2019}%
  \BibitemOpen
  \bibfield  {author} {\bibinfo {author} {\bibfnamefont {D.}~\bibnamefont
  {Marpaung}}, \bibinfo {author} {\bibfnamefont {J.}~\bibnamefont {Yao}}, \
  and\ \bibinfo {author} {\bibfnamefont {J.}~\bibnamefont {Capmany}},\ }\href
  {\doibase 10.1038/s41566-018-0310-5} {\bibfield  {journal} {\bibinfo
  {journal} {Nature Photonics}\ }\textbf {\bibinfo {volume} {13}},\ \bibinfo
  {pages} {80} (\bibinfo {year} {2019})}\BibitemShut {NoStop}%
\bibitem [{\citenamefont {Thévenaz}(2008)}]{thevenaz_slow_2008}%
  \BibitemOpen
  \bibfield  {author} {\bibinfo {author} {\bibfnamefont {L.}~\bibnamefont
  {Thévenaz}},\ }\href {\doibase 10.1038/nphoton.2008.147} {\bibfield
  {journal} {\bibinfo  {journal} {Nature Photonics}\ }\textbf {\bibinfo
  {volume} {2}},\ \bibinfo {pages} {474} (\bibinfo {year} {2008})}\BibitemShut
  {NoStop}%
\bibitem [{\citenamefont {Loh}\ \emph {et~al.}(2020)\citenamefont {Loh},
  \citenamefont {Stuart}, \citenamefont {Reens}, \citenamefont {Bruzewicz},
  \citenamefont {Braje}, \citenamefont {Chiaverini}, \citenamefont
  {Juodawlkis}, \citenamefont {Sage},\ and\ \citenamefont
  {McConnell}}]{loh_operation_2020}%
  \BibitemOpen
  \bibfield  {author} {\bibinfo {author} {\bibfnamefont {W.}~\bibnamefont
  {Loh}}, \bibinfo {author} {\bibfnamefont {J.}~\bibnamefont {Stuart}},
  \bibinfo {author} {\bibfnamefont {D.}~\bibnamefont {Reens}}, \bibinfo
  {author} {\bibfnamefont {C.~D.}\ \bibnamefont {Bruzewicz}}, \bibinfo {author}
  {\bibfnamefont {D.}~\bibnamefont {Braje}}, \bibinfo {author} {\bibfnamefont
  {J.}~\bibnamefont {Chiaverini}}, \bibinfo {author} {\bibfnamefont {P.~W.}\
  \bibnamefont {Juodawlkis}}, \bibinfo {author} {\bibfnamefont {J.~M.}\
  \bibnamefont {Sage}}, \ and\ \bibinfo {author} {\bibfnamefont
  {R.}~\bibnamefont {McConnell}},\ }\href {\doibase 10.1038/s41586-020-2981-6}
  {\bibfield  {journal} {\bibinfo  {journal} {Nature}\ }\textbf {\bibinfo
  {volume} {588}},\ \bibinfo {pages} {244} (\bibinfo {year}
  {2020})}\BibitemShut {NoStop}%
\bibitem [{\citenamefont {Yang}\ \emph {et~al.}(2020)\citenamefont {Yang},
  \citenamefont {Gyger},\ and\ \citenamefont {Thévenaz}}]{yang_intense_2020}%
  \BibitemOpen
  \bibfield  {author} {\bibinfo {author} {\bibfnamefont {F.}~\bibnamefont
  {Yang}}, \bibinfo {author} {\bibfnamefont {F.}~\bibnamefont {Gyger}}, \ and\
  \bibinfo {author} {\bibfnamefont {L.}~\bibnamefont {Thévenaz}},\ }\href
  {\doibase 10.1038/s41566-020-0676-z} {\bibfield  {journal} {\bibinfo
  {journal} {Nature Photonics}\ }\textbf {\bibinfo {volume} {14}},\ \bibinfo
  {pages} {700} (\bibinfo {year} {2020})}\BibitemShut {NoStop}%
\bibitem [{\citenamefont {Prevedel}\ \emph {et~al.}(2019)\citenamefont
  {Prevedel}, \citenamefont {Diz-Muñoz}, \citenamefont {Ruocco},\ and\
  \citenamefont {Antonacci}}]{prevedel_brillouin_2019}%
  \BibitemOpen
  \bibfield  {author} {\bibinfo {author} {\bibfnamefont {R.}~\bibnamefont
  {Prevedel}}, \bibinfo {author} {\bibfnamefont {A.}~\bibnamefont
  {Diz-Muñoz}}, \bibinfo {author} {\bibfnamefont {G.}~\bibnamefont {Ruocco}},
  \ and\ \bibinfo {author} {\bibfnamefont {G.}~\bibnamefont {Antonacci}},\
  }\href {\doibase 10.1038/s41592-019-0543-3} {\bibfield  {journal} {\bibinfo
  {journal} {Nature Methods}\ }\textbf {\bibinfo {volume} {16}},\ \bibinfo
  {pages} {969} (\bibinfo {year} {2019})}\BibitemShut {NoStop}%
\bibitem [{\citenamefont {Palombo}\ and\ \citenamefont
  {Fioretto}(2019)}]{palombo_brillouin_2019}%
  \BibitemOpen
  \bibfield  {author} {\bibinfo {author} {\bibfnamefont {F.}~\bibnamefont
  {Palombo}}\ and\ \bibinfo {author} {\bibfnamefont {D.}~\bibnamefont
  {Fioretto}},\ }\href {\doibase 10.1021/acs.chemrev.9b00019} {\bibfield
  {journal} {\bibinfo  {journal} {Chemical Reviews}\ }\textbf {\bibinfo
  {volume} {119}},\ \bibinfo {pages} {7833} (\bibinfo {year}
  {2019})}\BibitemShut {NoStop}%
\bibitem [{\citenamefont {Ippen}\ and\ \citenamefont
  {Stolen}(1972)}]{ippen_stimulated_1972}%
  \BibitemOpen
  \bibfield  {author} {\bibinfo {author} {\bibfnamefont {E.}~\bibnamefont
  {Ippen}}\ and\ \bibinfo {author} {\bibfnamefont {R.}~\bibnamefont {Stolen}},\
  }\href {\doibase 10.1063/1.1654249} {\bibfield  {journal} {\bibinfo
  {journal} {Applied Physics Letters}\ }\textbf {\bibinfo {volume} {21}},\
  \bibinfo {pages} {539} (\bibinfo {year} {1972})}\BibitemShut {NoStop}%
\bibitem [{\citenamefont {Dainese}\ \emph {et~al.}(2006)\citenamefont
  {Dainese}, \citenamefont {Russell}, \citenamefont {Joly}, \citenamefont
  {Knight}, \citenamefont {Wiederhecker}, \citenamefont {Fragnito},
  \citenamefont {Laude},\ and\ \citenamefont
  {Khelif}}]{dainese_stimulated_2006}%
  \BibitemOpen
  \bibfield  {author} {\bibinfo {author} {\bibfnamefont {P.}~\bibnamefont
  {Dainese}}, \bibinfo {author} {\bibfnamefont {P.~S.~J.}\ \bibnamefont
  {Russell}}, \bibinfo {author} {\bibfnamefont {N.}~\bibnamefont {Joly}},
  \bibinfo {author} {\bibfnamefont {J.~C.}\ \bibnamefont {Knight}}, \bibinfo
  {author} {\bibfnamefont {G.~S.}\ \bibnamefont {Wiederhecker}}, \bibinfo
  {author} {\bibfnamefont {H.~L.}\ \bibnamefont {Fragnito}}, \bibinfo {author}
  {\bibfnamefont {V.}~\bibnamefont {Laude}}, \ and\ \bibinfo {author}
  {\bibfnamefont {A.}~\bibnamefont {Khelif}},\ }\href {\doibase
  10.1038/nphys315} {\bibfield  {journal} {\bibinfo  {journal} {Nature
  Physics}\ }\textbf {\bibinfo {volume} {2}},\ \bibinfo {pages} {388} (\bibinfo
  {year} {2006})}\BibitemShut {NoStop}%
\bibitem [{\citenamefont {Beugnot}\ \emph {et~al.}(2014)\citenamefont
  {Beugnot}, \citenamefont {Lebrun}, \citenamefont {Pauliat}, \citenamefont
  {Maillotte}, \citenamefont {Laude},\ and\ \citenamefont
  {Sylvestre}}]{beugnot_brillouin_2014}%
  \BibitemOpen
  \bibfield  {author} {\bibinfo {author} {\bibfnamefont {J.-C.}\ \bibnamefont
  {Beugnot}}, \bibinfo {author} {\bibfnamefont {S.}~\bibnamefont {Lebrun}},
  \bibinfo {author} {\bibfnamefont {G.}~\bibnamefont {Pauliat}}, \bibinfo
  {author} {\bibfnamefont {H.}~\bibnamefont {Maillotte}}, \bibinfo {author}
  {\bibfnamefont {V.}~\bibnamefont {Laude}}, \ and\ \bibinfo {author}
  {\bibfnamefont {T.}~\bibnamefont {Sylvestre}},\ }\href {\doibase
  10.1038/ncomms6242} {\bibfield  {journal} {\bibinfo  {journal} {Nature
  Communications}\ }\textbf {\bibinfo {volume} {5}},\ \bibinfo {pages} {5242}
  (\bibinfo {year} {2014})}\BibitemShut {NoStop}%
\bibitem [{\citenamefont {Florez}\ \emph {et~al.}(2016)\citenamefont {Florez},
  \citenamefont {Jarschel}, \citenamefont {Espinel}, \citenamefont {Cordeiro},
  \citenamefont {Mayer~Alegre}, \citenamefont {Wiederhecker},\ and\
  \citenamefont {Dainese}}]{florez_brillouin_2016}%
  \BibitemOpen
  \bibfield  {author} {\bibinfo {author} {\bibfnamefont {O.}~\bibnamefont
  {Florez}}, \bibinfo {author} {\bibfnamefont {P.~F.}\ \bibnamefont
  {Jarschel}}, \bibinfo {author} {\bibfnamefont {Y.~a.~V.}\ \bibnamefont
  {Espinel}}, \bibinfo {author} {\bibfnamefont {C.~M.~B.}\ \bibnamefont
  {Cordeiro}}, \bibinfo {author} {\bibfnamefont {T.~P.}\ \bibnamefont
  {Mayer~Alegre}}, \bibinfo {author} {\bibfnamefont {G.~S.}\ \bibnamefont
  {Wiederhecker}}, \ and\ \bibinfo {author} {\bibfnamefont {P.}~\bibnamefont
  {Dainese}},\ }\href {\doibase 10.1038/ncomms11759} {\bibfield  {journal}
  {\bibinfo  {journal} {Nature Communications}\ }\textbf {\bibinfo {volume}
  {7}},\ \bibinfo {pages} {11759} (\bibinfo {year} {2016})}\BibitemShut
  {NoStop}%
\bibitem [{\citenamefont {Grudinin}\ \emph {et~al.}(2009)\citenamefont
  {Grudinin}, \citenamefont {Matsko},\ and\ \citenamefont
  {Maleki}}]{grudinin_brillouin_2009}%
  \BibitemOpen
  \bibfield  {author} {\bibinfo {author} {\bibfnamefont {I.~S.}\ \bibnamefont
  {Grudinin}}, \bibinfo {author} {\bibfnamefont {A.~B.}\ \bibnamefont
  {Matsko}}, \ and\ \bibinfo {author} {\bibfnamefont {L.}~\bibnamefont
  {Maleki}},\ }\href {\doibase 10.1103/PhysRevLett.102.043902} {\bibfield
  {journal} {\bibinfo  {journal} {Physical Review Letters}\ }\textbf {\bibinfo
  {volume} {102}},\ \bibinfo {pages} {043902} (\bibinfo {year}
  {2009})}\BibitemShut {NoStop}%
\bibitem [{\citenamefont {Tomes}\ and\ \citenamefont
  {Carmon}(2009)}]{tomes_photonic_2009}%
  \BibitemOpen
  \bibfield  {author} {\bibinfo {author} {\bibfnamefont {M.}~\bibnamefont
  {Tomes}}\ and\ \bibinfo {author} {\bibfnamefont {T.}~\bibnamefont {Carmon}},\
  }\href {\doibase 10.1103/PhysRevLett.102.113601} {\bibfield  {journal}
  {\bibinfo  {journal} {Physical Review Letters}\ }\textbf {\bibinfo {volume}
  {102}},\ \bibinfo {pages} {113601} (\bibinfo {year} {2009})}\BibitemShut
  {NoStop}%
\bibitem [{\citenamefont {Lee}\ \emph {et~al.}(2012)\citenamefont {Lee},
  \citenamefont {Chen}, \citenamefont {Li}, \citenamefont {Yang}, \citenamefont
  {Jeon}, \citenamefont {Painter},\ and\ \citenamefont
  {Vahala}}]{lee_chemically_2012}%
  \BibitemOpen
  \bibfield  {author} {\bibinfo {author} {\bibfnamefont {H.}~\bibnamefont
  {Lee}}, \bibinfo {author} {\bibfnamefont {T.}~\bibnamefont {Chen}}, \bibinfo
  {author} {\bibfnamefont {J.}~\bibnamefont {Li}}, \bibinfo {author}
  {\bibfnamefont {K.~Y.}\ \bibnamefont {Yang}}, \bibinfo {author}
  {\bibfnamefont {S.}~\bibnamefont {Jeon}}, \bibinfo {author} {\bibfnamefont
  {O.}~\bibnamefont {Painter}}, \ and\ \bibinfo {author} {\bibfnamefont
  {K.~J.}\ \bibnamefont {Vahala}},\ }\href {\doibase 10.1038/nphoton.2012.109}
  {\bibfield  {journal} {\bibinfo  {journal} {Nature Photonics}\ }\textbf
  {\bibinfo {volume} {6}},\ \bibinfo {pages} {369} (\bibinfo {year}
  {2012})}\BibitemShut {NoStop}%
\bibitem [{\citenamefont {Kim}\ \emph {et~al.}(2015)\citenamefont {Kim},
  \citenamefont {Kuzyk}, \citenamefont {Han}, \citenamefont {Wang},\ and\
  \citenamefont {Bahl}}]{kim_non-reciprocal_2015}%
  \BibitemOpen
  \bibfield  {author} {\bibinfo {author} {\bibfnamefont {J.}~\bibnamefont
  {Kim}}, \bibinfo {author} {\bibfnamefont {M.~C.}\ \bibnamefont {Kuzyk}},
  \bibinfo {author} {\bibfnamefont {K.}~\bibnamefont {Han}}, \bibinfo {author}
  {\bibfnamefont {H.}~\bibnamefont {Wang}}, \ and\ \bibinfo {author}
  {\bibfnamefont {G.}~\bibnamefont {Bahl}},\ }\href {\doibase
  10.1038/nphys3236} {\bibfield  {journal} {\bibinfo  {journal} {Nature
  Physics}\ }\textbf {\bibinfo {volume} {11}},\ \bibinfo {pages} {275}
  (\bibinfo {year} {2015})}\BibitemShut {NoStop}%
\bibitem [{\citenamefont {Pant}\ \emph {et~al.}(2011)\citenamefont {Pant},
  \citenamefont {Poulton}, \citenamefont {Choi}, \citenamefont {Mcfarlane},
  \citenamefont {Hile}, \citenamefont {Li}, \citenamefont {Th{\'e}venaz},
  \citenamefont {Luther-Davies}, \citenamefont {Madden},\ and\ \citenamefont
  {Eggleton}}]{pant_-chip_2011}%
  \BibitemOpen
  \bibfield  {author} {\bibinfo {author} {\bibfnamefont {R.}~\bibnamefont
  {Pant}}, \bibinfo {author} {\bibfnamefont {C.~G.}\ \bibnamefont {Poulton}},
  \bibinfo {author} {\bibfnamefont {D.-Y.}\ \bibnamefont {Choi}}, \bibinfo
  {author} {\bibfnamefont {H.}~\bibnamefont {Mcfarlane}}, \bibinfo {author}
  {\bibfnamefont {S.}~\bibnamefont {Hile}}, \bibinfo {author} {\bibfnamefont
  {E.}~\bibnamefont {Li}}, \bibinfo {author} {\bibfnamefont {L.}~\bibnamefont
  {Th{\'e}venaz}}, \bibinfo {author} {\bibfnamefont {B.}~\bibnamefont
  {Luther-Davies}}, \bibinfo {author} {\bibfnamefont {S.~J.}\ \bibnamefont
  {Madden}}, \ and\ \bibinfo {author} {\bibfnamefont {B.~J.}\ \bibnamefont
  {Eggleton}},\ }\href {\doibase 10.1364/OE.19.008285} {\bibfield  {journal}
  {\bibinfo  {journal} {Optics Express}\ }\textbf {\bibinfo {volume} {19}},\
  \bibinfo {pages} {8285} (\bibinfo {year} {2011})}\BibitemShut {NoStop}%
\bibitem [{\citenamefont {Shin}\ \emph {et~al.}(2013)\citenamefont {Shin},
  \citenamefont {Qiu}, \citenamefont {Jarecki}, \citenamefont {Cox},
  \citenamefont {Olsson~Iii}, \citenamefont {Starbuck}, \citenamefont {Wang},\
  and\ \citenamefont {Rakich}}]{shin_tailorable_2013}%
  \BibitemOpen
  \bibfield  {author} {\bibinfo {author} {\bibfnamefont {H.}~\bibnamefont
  {Shin}}, \bibinfo {author} {\bibfnamefont {W.}~\bibnamefont {Qiu}}, \bibinfo
  {author} {\bibfnamefont {R.}~\bibnamefont {Jarecki}}, \bibinfo {author}
  {\bibfnamefont {J.~A.}\ \bibnamefont {Cox}}, \bibinfo {author} {\bibfnamefont
  {R.~H.}\ \bibnamefont {Olsson~Iii}}, \bibinfo {author} {\bibfnamefont
  {A.}~\bibnamefont {Starbuck}}, \bibinfo {author} {\bibfnamefont
  {Z.}~\bibnamefont {Wang}}, \ and\ \bibinfo {author} {\bibfnamefont {P.~T.}\
  \bibnamefont {Rakich}},\ }\href {\doibase 10.1038/ncomms2943} {\bibfield
  {journal} {\bibinfo  {journal} {Nature Communications}\ }\textbf {\bibinfo
  {volume} {4}},\ \bibinfo {pages} {1944} (\bibinfo {year} {2013})}\BibitemShut
  {NoStop}%
\bibitem [{\citenamefont {Van~Laer}\ \emph {et~al.}(2015)\citenamefont
  {Van~Laer}, \citenamefont {Kuyken}, \citenamefont {Van~Thourhout},\ and\
  \citenamefont {Baets}}]{van_laer_interaction_2015}%
  \BibitemOpen
  \bibfield  {author} {\bibinfo {author} {\bibfnamefont {R.}~\bibnamefont
  {Van~Laer}}, \bibinfo {author} {\bibfnamefont {B.}~\bibnamefont {Kuyken}},
  \bibinfo {author} {\bibfnamefont {D.}~\bibnamefont {Van~Thourhout}}, \ and\
  \bibinfo {author} {\bibfnamefont {R.}~\bibnamefont {Baets}},\ }\href
  {\doibase 10.1038/nphoton.2015.11} {\bibfield  {journal} {\bibinfo  {journal}
  {Nature Photonics}\ }\textbf {\bibinfo {volume} {9}},\ \bibinfo {pages} {199}
  (\bibinfo {year} {2015})}\BibitemShut {NoStop}%
\bibitem [{\citenamefont {Yang}\ \emph {et~al.}(2018)\citenamefont {Yang},
  \citenamefont {Oh}, \citenamefont {Lee}, \citenamefont {Yang}, \citenamefont
  {Yi}, \citenamefont {Shen}, \citenamefont {Wang},\ and\ \citenamefont
  {Vahala}}]{yang_bridging_2018}%
  \BibitemOpen
  \bibfield  {author} {\bibinfo {author} {\bibfnamefont {K.~Y.}\ \bibnamefont
  {Yang}}, \bibinfo {author} {\bibfnamefont {D.~Y.}\ \bibnamefont {Oh}},
  \bibinfo {author} {\bibfnamefont {S.~H.}\ \bibnamefont {Lee}}, \bibinfo
  {author} {\bibfnamefont {Q.-F.}\ \bibnamefont {Yang}}, \bibinfo {author}
  {\bibfnamefont {X.}~\bibnamefont {Yi}}, \bibinfo {author} {\bibfnamefont
  {B.}~\bibnamefont {Shen}}, \bibinfo {author} {\bibfnamefont {H.}~\bibnamefont
  {Wang}}, \ and\ \bibinfo {author} {\bibfnamefont {K.}~\bibnamefont
  {Vahala}},\ }\href {\doibase 10.1038/s41566-018-0132-5} {\bibfield  {journal}
  {\bibinfo  {journal} {Nature Photonics}\ }\textbf {\bibinfo {volume} {12}},\
  \bibinfo {pages} {297} (\bibinfo {year} {2018})}\BibitemShut {NoStop}%
\bibitem [{\citenamefont {Otterstrom}\ \emph {et~al.}(2018)\citenamefont
  {Otterstrom}, \citenamefont {Behunin}, \citenamefont {Kittlaus},
  \citenamefont {Wang},\ and\ \citenamefont
  {Rakich}}]{otterstrom_silicon_2018}%
  \BibitemOpen
  \bibfield  {author} {\bibinfo {author} {\bibfnamefont {N.~T.}\ \bibnamefont
  {Otterstrom}}, \bibinfo {author} {\bibfnamefont {R.~O.}\ \bibnamefont
  {Behunin}}, \bibinfo {author} {\bibfnamefont {E.~A.}\ \bibnamefont
  {Kittlaus}}, \bibinfo {author} {\bibfnamefont {Z.}~\bibnamefont {Wang}}, \
  and\ \bibinfo {author} {\bibfnamefont {P.~T.}\ \bibnamefont {Rakich}},\
  }\href {\doibase 10.1126/science.aar6113} {\bibfield  {journal} {\bibinfo
  {journal} {Science}\ }\textbf {\bibinfo {volume} {360}},\ \bibinfo {pages}
  {1113} (\bibinfo {year} {2018})}\BibitemShut {NoStop}%
\bibitem [{\citenamefont {Gundavarapu}\ \emph {et~al.}(2019)\citenamefont
  {Gundavarapu}, \citenamefont {Brodnik}, \citenamefont {Puckett},
  \citenamefont {Huffman}, \citenamefont {Bose}, \citenamefont {Behunin},
  \citenamefont {Wu}, \citenamefont {Qiu}, \citenamefont {Pinho}, \citenamefont
  {Chauhan}, \citenamefont {Nohava}, \citenamefont {Rakich}, \citenamefont
  {Nelson}, \citenamefont {Salit},\ and\ \citenamefont
  {Blumenthal}}]{gundavarapu_sub-hertz_2019}%
  \BibitemOpen
  \bibfield  {author} {\bibinfo {author} {\bibfnamefont {S.}~\bibnamefont
  {Gundavarapu}}, \bibinfo {author} {\bibfnamefont {G.~M.}\ \bibnamefont
  {Brodnik}}, \bibinfo {author} {\bibfnamefont {M.}~\bibnamefont {Puckett}},
  \bibinfo {author} {\bibfnamefont {T.}~\bibnamefont {Huffman}}, \bibinfo
  {author} {\bibfnamefont {D.}~\bibnamefont {Bose}}, \bibinfo {author}
  {\bibfnamefont {R.}~\bibnamefont {Behunin}}, \bibinfo {author} {\bibfnamefont
  {J.}~\bibnamefont {Wu}}, \bibinfo {author} {\bibfnamefont {T.}~\bibnamefont
  {Qiu}}, \bibinfo {author} {\bibfnamefont {C.}~\bibnamefont {Pinho}}, \bibinfo
  {author} {\bibfnamefont {N.}~\bibnamefont {Chauhan}}, \bibinfo {author}
  {\bibfnamefont {J.}~\bibnamefont {Nohava}}, \bibinfo {author} {\bibfnamefont
  {P.~T.}\ \bibnamefont {Rakich}}, \bibinfo {author} {\bibfnamefont {K.~D.}\
  \bibnamefont {Nelson}}, \bibinfo {author} {\bibfnamefont {M.}~\bibnamefont
  {Salit}}, \ and\ \bibinfo {author} {\bibfnamefont {D.~J.}\ \bibnamefont
  {Blumenthal}},\ }\href {\doibase 10.1038/s41566-018-0313-2} {\bibfield
  {journal} {\bibinfo  {journal} {Nature Photonics}\ }\textbf {\bibinfo
  {volume} {13}},\ \bibinfo {pages} {60} (\bibinfo {year} {2019})}\BibitemShut
  {NoStop}%
\bibitem [{\citenamefont {Gyger}\ \emph {et~al.}(2020)\citenamefont {Gyger},
  \citenamefont {Liu}, \citenamefont {Yang}, \citenamefont {He}, \citenamefont
  {Raja}, \citenamefont {Wang}, \citenamefont {Bhave}, \citenamefont
  {Kippenberg},\ and\ \citenamefont {Thévenaz}}]{gyger_observation_2020}%
  \BibitemOpen
  \bibfield  {author} {\bibinfo {author} {\bibfnamefont {F.}~\bibnamefont
  {Gyger}}, \bibinfo {author} {\bibfnamefont {J.}~\bibnamefont {Liu}}, \bibinfo
  {author} {\bibfnamefont {F.}~\bibnamefont {Yang}}, \bibinfo {author}
  {\bibfnamefont {J.}~\bibnamefont {He}}, \bibinfo {author} {\bibfnamefont
  {A.~S.}\ \bibnamefont {Raja}}, \bibinfo {author} {\bibfnamefont {R.~N.}\
  \bibnamefont {Wang}}, \bibinfo {author} {\bibfnamefont {S.~A.}\ \bibnamefont
  {Bhave}}, \bibinfo {author} {\bibfnamefont {T.~J.}\ \bibnamefont
  {Kippenberg}}, \ and\ \bibinfo {author} {\bibfnamefont {L.}~\bibnamefont
  {Thévenaz}},\ }\href {\doibase 10.1103/PhysRevLett.124.013902} {\bibfield
  {journal} {\bibinfo  {journal} {Physical Review Letters}\ }\textbf {\bibinfo
  {volume} {124}},\ \bibinfo {pages} {013902} (\bibinfo {year}
  {2020})}\BibitemShut {NoStop}%
\bibitem [{\citenamefont {Rakich}\ \emph {et~al.}(2012)\citenamefont {Rakich},
  \citenamefont {Reinke}, \citenamefont {Camacho}, \citenamefont {Davids},\
  and\ \citenamefont {Wang}}]{rakich_giant_2012}%
  \BibitemOpen
  \bibfield  {author} {\bibinfo {author} {\bibfnamefont {P.~T.}\ \bibnamefont
  {Rakich}}, \bibinfo {author} {\bibfnamefont {C.}~\bibnamefont {Reinke}},
  \bibinfo {author} {\bibfnamefont {R.}~\bibnamefont {Camacho}}, \bibinfo
  {author} {\bibfnamefont {P.}~\bibnamefont {Davids}}, \ and\ \bibinfo {author}
  {\bibfnamefont {Z.}~\bibnamefont {Wang}},\ }\href {\doibase
  10.1103/PhysRevX.2.011008} {\bibfield  {journal} {\bibinfo  {journal}
  {Physical Review X}\ }\textbf {\bibinfo {volume} {2}},\ \bibinfo {pages}
  {011008} (\bibinfo {year} {2012})}\BibitemShut {NoStop}%
\bibitem [{\citenamefont {Scarcelli}\ and\ \citenamefont
  {Yun}(2008)}]{scarcelli_confocal_2008}%
  \BibitemOpen
  \bibfield  {author} {\bibinfo {author} {\bibfnamefont {G.}~\bibnamefont
  {Scarcelli}}\ and\ \bibinfo {author} {\bibfnamefont {S.~H.}\ \bibnamefont
  {Yun}},\ }\href {\doibase 10.1038/nphoton.2007.250} {\bibfield  {journal}
  {\bibinfo  {journal} {Nature Photonics}\ }\textbf {\bibinfo {volume} {2}},\
  \bibinfo {pages} {39} (\bibinfo {year} {2008})}\BibitemShut {NoStop}%
\bibitem [{\citenamefont {Bailey}\ \emph {et~al.}(2020)\citenamefont {Bailey},
  \citenamefont {Alunni-Cardinali}, \citenamefont {Correa}, \citenamefont
  {Caponi}, \citenamefont {Holsgrove}, \citenamefont {Barr}, \citenamefont
  {Stone}, \citenamefont {Winlove}, \citenamefont {Fioretto},\ and\
  \citenamefont {Palombo}}]{bailey_viscoelastic_2020}%
  \BibitemOpen
  \bibfield  {author} {\bibinfo {author} {\bibfnamefont {M.}~\bibnamefont
  {Bailey}}, \bibinfo {author} {\bibfnamefont {M.}~\bibnamefont
  {Alunni-Cardinali}}, \bibinfo {author} {\bibfnamefont {N.}~\bibnamefont
  {Correa}}, \bibinfo {author} {\bibfnamefont {S.}~\bibnamefont {Caponi}},
  \bibinfo {author} {\bibfnamefont {T.}~\bibnamefont {Holsgrove}}, \bibinfo
  {author} {\bibfnamefont {H.}~\bibnamefont {Barr}}, \bibinfo {author}
  {\bibfnamefont {N.}~\bibnamefont {Stone}}, \bibinfo {author} {\bibfnamefont
  {C.~P.}\ \bibnamefont {Winlove}}, \bibinfo {author} {\bibfnamefont
  {D.}~\bibnamefont {Fioretto}}, \ and\ \bibinfo {author} {\bibfnamefont
  {F.}~\bibnamefont {Palombo}},\ }\href {\doibase 10.1126/sciadv.abc1937}
  {\bibfield  {journal} {\bibinfo  {journal} {Science Advances}\ }\textbf
  {\bibinfo {volume} {6}},\ \bibinfo {pages} {eabc1937} (\bibinfo {year}
  {2020})}\BibitemShut {NoStop}%
\bibitem [{\citenamefont {Coucheron}\ \emph {et~al.}(2019)\citenamefont
  {Coucheron}, \citenamefont {Helle}, \citenamefont {Oie}, \citenamefont
  {Tinguely},\ and\ \citenamefont
  {Ahluwalia}}]{coucheron_high-throughput_2019}%
  \BibitemOpen
  \bibfield  {author} {\bibinfo {author} {\bibfnamefont {D.~A.}\ \bibnamefont
  {Coucheron}}, \bibinfo {author} {\bibfnamefont {O.~I.}\ \bibnamefont
  {Helle}}, \bibinfo {author} {\bibfnamefont {C.~I.}\ \bibnamefont {Oie}},
  \bibinfo {author} {\bibfnamefont {J.-C.}\ \bibnamefont {Tinguely}}, \ and\
  \bibinfo {author} {\bibfnamefont {B.~S.}\ \bibnamefont {Ahluwalia}},\ }\href
  {\doibase 10.3791/60378} {\bibfield  {journal} {\bibinfo  {journal} {JoVE
  (Journal of Visualized Experiments)}\ ,\ \bibinfo {pages} {e60378}} (\bibinfo
  {year} {2019})}\BibitemShut {NoStop}%
\bibitem [{\citenamefont {Diekmann}\ \emph {et~al.}(2017)\citenamefont
  {Diekmann}, \citenamefont {Helle}, \citenamefont {Oie}, \citenamefont
  {McCourt}, \citenamefont {Huser}, \citenamefont {Schuttpelz},\ and\
  \citenamefont {Ahluwalia}}]{diekmann_chip-based_2017}%
  \BibitemOpen
  \bibfield  {author} {\bibinfo {author} {\bibfnamefont {R.}~\bibnamefont
  {Diekmann}}, \bibinfo {author} {\bibfnamefont {O.~I.}\ \bibnamefont {Helle}},
  \bibinfo {author} {\bibfnamefont {C.~I.}\ \bibnamefont {Oie}}, \bibinfo
  {author} {\bibfnamefont {P.}~\bibnamefont {McCourt}}, \bibinfo {author}
  {\bibfnamefont {T.~R.}\ \bibnamefont {Huser}}, \bibinfo {author}
  {\bibfnamefont {M.}~\bibnamefont {Schuttpelz}}, \ and\ \bibinfo {author}
  {\bibfnamefont {B.~S.}\ \bibnamefont {Ahluwalia}},\ }\href {\doibase
  10.1038/nphoton.2017.55} {\bibfield  {journal} {\bibinfo  {journal} {Nature
  Photonics}\ }\textbf {\bibinfo {volume} {11}},\ \bibinfo {pages} {322}
  (\bibinfo {year} {2017})}\BibitemShut {NoStop}%
\bibitem [{\citenamefont {Helle}\ \emph {et~al.}(2020)\citenamefont {Helle},
  \citenamefont {Dullo}, \citenamefont {Lahrberg}, \citenamefont {Tinguely},
  \citenamefont {Helleso},\ and\ \citenamefont
  {Ahluwalia}}]{helle_structured_2020}%
  \BibitemOpen
  \bibfield  {author} {\bibinfo {author} {\bibfnamefont {O.~I.}\ \bibnamefont
  {Helle}}, \bibinfo {author} {\bibfnamefont {F.~T.}\ \bibnamefont {Dullo}},
  \bibinfo {author} {\bibfnamefont {M.}~\bibnamefont {Lahrberg}}, \bibinfo
  {author} {\bibfnamefont {J.-C.}\ \bibnamefont {Tinguely}}, \bibinfo {author}
  {\bibfnamefont {O.~G.}\ \bibnamefont {Helleso}}, \ and\ \bibinfo {author}
  {\bibfnamefont {B.~S.}\ \bibnamefont {Ahluwalia}},\ }\href {\doibase
  10.1038/s41566-020-0620-2} {\bibfield  {journal} {\bibinfo  {journal} {Nature
  Photonics}\ }\textbf {\bibinfo {volume} {14}},\ \bibinfo {pages} {431}
  (\bibinfo {year} {2020})}\BibitemShut {NoStop}%
\bibitem [{\citenamefont {Boyd}(2008)}]{boyd_nonlinear_2008}%
  \BibitemOpen
  \bibfield  {author} {\bibinfo {author} {\bibfnamefont {R.~W.}\ \bibnamefont
  {Boyd}},\ }\href
  {https://www.elsevier.com/books/nonlinear-optics/boyd/978-0-12-369470-6}
  {\emph {\bibinfo {title} {Nonlinear Optics - 3rd Edition}}}\ (\bibinfo
  {publisher} {Academic Press},\ \bibinfo {year} {2008})\BibitemShut {NoStop}%
\bibitem [{\citenamefont {Godet}\ \emph {et~al.}(2017)\citenamefont {Godet},
  \citenamefont {Ndao}, \citenamefont {Sylvestre}, \citenamefont {Pecheur},
  \citenamefont {Lebrun}, \citenamefont {Pauliat}, \citenamefont {Beugnot},\
  and\ \citenamefont {Huy}}]{godet_brillouin_2017}%
  \BibitemOpen
  \bibfield  {author} {\bibinfo {author} {\bibfnamefont {A.}~\bibnamefont
  {Godet}}, \bibinfo {author} {\bibfnamefont {A.}~\bibnamefont {Ndao}},
  \bibinfo {author} {\bibfnamefont {T.}~\bibnamefont {Sylvestre}}, \bibinfo
  {author} {\bibfnamefont {V.}~\bibnamefont {Pecheur}}, \bibinfo {author}
  {\bibfnamefont {S.}~\bibnamefont {Lebrun}}, \bibinfo {author} {\bibfnamefont
  {G.}~\bibnamefont {Pauliat}}, \bibinfo {author} {\bibfnamefont {J.-C.}\
  \bibnamefont {Beugnot}}, \ and\ \bibinfo {author} {\bibfnamefont {K.~P.}\
  \bibnamefont {Huy}},\ }\href {\doibase 10.1364/OPTICA.4.001232} {\bibfield
  {journal} {\bibinfo  {journal} {Optica}\ }\textbf {\bibinfo {volume} {4}},\
  \bibinfo {pages} {1232} (\bibinfo {year} {2017})}\BibitemShut {NoStop}%
\bibitem [{\citenamefont {Godet}\ \emph {et~al.}(2019)\citenamefont {Godet},
  \citenamefont {Sylvestre}, \citenamefont {Pêcheur}, \citenamefont
  {Chrétien}, \citenamefont {Beugnot},\ and\ \citenamefont
  {Phan~Huy}}]{godet_nonlinear_2019}%
  \BibitemOpen
  \bibfield  {author} {\bibinfo {author} {\bibfnamefont {A.}~\bibnamefont
  {Godet}}, \bibinfo {author} {\bibfnamefont {T.}~\bibnamefont {Sylvestre}},
  \bibinfo {author} {\bibfnamefont {V.}~\bibnamefont {Pêcheur}}, \bibinfo
  {author} {\bibfnamefont {J.}~\bibnamefont {Chrétien}}, \bibinfo {author}
  {\bibfnamefont {J.-C.}\ \bibnamefont {Beugnot}}, \ and\ \bibinfo {author}
  {\bibfnamefont {K.}~\bibnamefont {Phan~Huy}},\ }\href {\doibase
  10.1063/1.5103239} {\bibfield  {journal} {\bibinfo  {journal} {APL
  Photonics}\ }\textbf {\bibinfo {volume} {4}},\ \bibinfo {pages} {080804}
  (\bibinfo {year} {2019})}\BibitemShut {NoStop}%
\bibitem [{\citenamefont {Boyd}\ \emph {et~al.}(1990)\citenamefont {Boyd},
  \citenamefont {Rzaewski},\ and\ \citenamefont {Narum}}]{boyd_noise_1990}%
  \BibitemOpen
  \bibfield  {author} {\bibinfo {author} {\bibfnamefont {R.~W.}\ \bibnamefont
  {Boyd}}, \bibinfo {author} {\bibfnamefont {K.}~\bibnamefont {Rzaewski}}, \
  and\ \bibinfo {author} {\bibfnamefont {P.}~\bibnamefont {Narum}},\ }\href
  {\doibase 10.1103/PhysRevA.42.5514} {\bibfield  {journal} {\bibinfo
  {journal} {Physical Review A}\ }\textbf {\bibinfo {volume} {42}},\ \bibinfo
  {pages} {5514} (\bibinfo {year} {1990})}\BibitemShut {NoStop}%
\bibitem [{\citenamefont {Motil}\ \emph {et~al.}(2016)\citenamefont {Motil},
  \citenamefont {Bergman},\ and\ \citenamefont {Tur}}]{motil_invited_2016}%
  \BibitemOpen
  \bibfield  {author} {\bibinfo {author} {\bibfnamefont {A.}~\bibnamefont
  {Motil}}, \bibinfo {author} {\bibfnamefont {A.}~\bibnamefont {Bergman}}, \
  and\ \bibinfo {author} {\bibfnamefont {M.}~\bibnamefont {Tur}},\ }\href
  {\doibase 10.1016/j.optlastec.2015.09.013} {\bibfield  {journal} {\bibinfo
  {journal} {Optics \& Laser Technology}\ }\textbf {\bibinfo {volume} {78}},\
  \bibinfo {pages} {81} (\bibinfo {year} {2016})}\BibitemShut {NoStop}%
\bibitem [{\citenamefont {Poulton}\ \emph {et~al.}(2013)\citenamefont
  {Poulton}, \citenamefont {Pant},\ and\ \citenamefont
  {Eggleton}}]{poulton2013acoustic}%
  \BibitemOpen
  \bibfield  {author} {\bibinfo {author} {\bibfnamefont {C.~G.}\ \bibnamefont
  {Poulton}}, \bibinfo {author} {\bibfnamefont {R.}~\bibnamefont {Pant}}, \
  and\ \bibinfo {author} {\bibfnamefont {B.~J.}\ \bibnamefont {Eggleton}},\
  }\href {\doibase 10.1364/JOSAB.30.002657} {\bibfield  {journal} {\bibinfo
  {journal} {Journal of the Optical Society of America B}\ }\textbf {\bibinfo
  {volume} {30}},\ \bibinfo {pages} {2657} (\bibinfo {year}
  {2013})}\BibitemShut {NoStop}%
\bibitem [{\citenamefont {Eden}\ and\ \citenamefont
  {Swinney}(1974)}]{eden_optical_1974}%
  \BibitemOpen
  \bibfield  {author} {\bibinfo {author} {\bibfnamefont {D.}~\bibnamefont
  {Eden}}\ and\ \bibinfo {author} {\bibfnamefont {H.~L.}\ \bibnamefont
  {Swinney}},\ }\href {\doibase 10.1016/0030-4018(74)90052-2} {\bibfield
  {journal} {\bibinfo  {journal} {Optics Communications}\ }\textbf {\bibinfo
  {volume} {10}},\ \bibinfo {pages} {191} (\bibinfo {year} {1974})}\BibitemShut
  {NoStop}%
\bibitem [{\citenamefont {Almeida}\ \emph {et~al.}(2004)\citenamefont
  {Almeida}, \citenamefont {Xu}, \citenamefont {Barrios},\ and\ \citenamefont
  {Lipson}}]{almeida_guiding_2004}%
  \BibitemOpen
  \bibfield  {author} {\bibinfo {author} {\bibfnamefont {V.~R.}\ \bibnamefont
  {Almeida}}, \bibinfo {author} {\bibfnamefont {Q.}~\bibnamefont {Xu}},
  \bibinfo {author} {\bibfnamefont {C.~A.}\ \bibnamefont {Barrios}}, \ and\
  \bibinfo {author} {\bibfnamefont {M.}~\bibnamefont {Lipson}},\ }\href
  {\doibase 10.1364/OL.29.001209} {\bibfield  {journal} {\bibinfo  {journal}
  {Optics Letters}\ }\textbf {\bibinfo {volume} {29}},\ \bibinfo {pages} {1209}
  (\bibinfo {year} {2004})}\BibitemShut {NoStop}%
\bibitem [{\citenamefont {Laude}\ \emph {et~al.}(2018)\citenamefont {Laude},
  \citenamefont {Korotyaeva},\ and\ \citenamefont {Beugnot}}]{Laude_2018}%
  \BibitemOpen
  \bibfield  {author} {\bibinfo {author} {\bibfnamefont {V.}~\bibnamefont
  {Laude}}, \bibinfo {author} {\bibfnamefont {M.~E.}\ \bibnamefont
  {Korotyaeva}}, \ and\ \bibinfo {author} {\bibfnamefont {J.-C.}\ \bibnamefont
  {Beugnot}},\ }\href {\doibase 10.1364/AO.57.000C77} {\bibfield  {journal}
  {\bibinfo  {journal} {Appl. Opt.}\ }\textbf {\bibinfo {volume} {57}},\
  \bibinfo {pages} {C77} (\bibinfo {year} {2018})}\BibitemShut {NoStop}%
\bibitem [{\citenamefont {Beugnot}\ and\ \citenamefont
  {Laude}(2012)}]{Beugnot_2012}%
  \BibitemOpen
  \bibfield  {author} {\bibinfo {author} {\bibfnamefont {J.-C.}\ \bibnamefont
  {Beugnot}}\ and\ \bibinfo {author} {\bibfnamefont {V.}~\bibnamefont
  {Laude}},\ }\href {\doibase 10.1103/PhysRevB.86.224304} {\bibfield  {journal}
  {\bibinfo  {journal} {Phys. Rev. B}\ }\textbf {\bibinfo {volume} {86}},\
  \bibinfo {pages} {224304} (\bibinfo {year} {2012})}\BibitemShut {NoStop}%
\end{thebibliography}%


\begin{thebibliography}{1}
\expandafter\ifx\csname url\endcsname\relax
  \def\url#1{\texttt{#1}}\fi
\expandafter\ifx\csname urlprefix\endcsname\relax\def\urlprefix{URL }\fi
\providecommand{\bibinfo}[2]{#2}
\providecommand{\eprint}[2][]{\url{#2}}

\bibitem{boyd1990noise}
\bibinfo{author}{Boyd, R.~W.}, \bibinfo{author}{Rzazewski, K.} \&
  \bibinfo{author}{Narum, P.}
\newblock \bibinfo{title}{Noise initiation of stimulated {Brillouin}
  scattering}.
\newblock \emph{\bibinfo{journal}{Physical Review A}}
  \textbf{\bibinfo{volume}{42}}, \bibinfo{pages}{5514} (\bibinfo{year}{1990}).
\newblock
  \urlprefix\url{https://journals.aps.org/pra/abstract/10.1103/PhysRevA.42.5514}.

\bibitem{renninger_forward_2016}
\bibinfo{author}{Renninger, W.~H.} \emph{et~al.}
\newblock \bibinfo{title}{Forward {Brillouin} scattering in hollow-core
  photonic bandgap fibers}.
\newblock \emph{\bibinfo{journal}{New Journal of Physics}}
  \textbf{\bibinfo{volume}{18}}, \bibinfo{pages}{025008}
  (\bibinfo{year}{2016}).
\newblock \urlprefix\url{https://doi.org/10.1088/1367-2630/18/2/025008}.

\bibitem{renninger_guided-wave_2016}
\bibinfo{author}{Renninger, W.~H.}, \bibinfo{author}{Behunin, R.~O.} \&
  \bibinfo{author}{Rakich, P.~T.}
\newblock \bibinfo{title}{Guided-wave {Brillouin} scattering in air}.
\newblock \emph{\bibinfo{journal}{Optica}} \textbf{\bibinfo{volume}{3}},
  \bibinfo{pages}{1316--1319} (\bibinfo{year}{2016}).
\newblock
  \urlprefix\url{https://www.osapublishing.org/optica/abstract.cfm?uri=optica-3-12-1316}.

\bibitem{motil_invited_2016}
\bibinfo{author}{Motil, A.}, \bibinfo{author}{Bergman, A.} \&
  \bibinfo{author}{Tur, M.}
\newblock \bibinfo{title}{[{INVITED}] {State} of the art of {Brillouin}
  fiber-optic distributed sensing}.
\newblock \emph{\bibinfo{journal}{Optics \& Laser Technology}}
  \textbf{\bibinfo{volume}{78}}, \bibinfo{pages}{81--103}
  (\bibinfo{year}{2016}).
\newblock
  \urlprefix\url{https://www.sciencedirect.com/science/article/pii/S0030399215002571}.

\bibitem{Hoffman_2014}
\bibinfo{author}{Hoffman, J.~E.} \emph{et~al.}
\newblock \bibinfo{title}{Ultrahigh transmission optical nanofibers}.
\newblock \emph{\bibinfo{journal}{AIP Advances}} \textbf{\bibinfo{volume}{4}},
  \bibinfo{pages}{067124} (\bibinfo{year}{2014}).
\newblock \urlprefix\url{https://doi.org/10.1063/1.4879799}.

\bibitem{Chow_2018}
\bibinfo{author}{Chow, D.~M.} \emph{et~al.}
\newblock \bibinfo{title}{Local activation of surface and hybrid acoustic waves
  in optical microwires}.
\newblock \emph{\bibinfo{journal}{Opt. Lett.}} \textbf{\bibinfo{volume}{43}},
  \bibinfo{pages}{1487--1490} (\bibinfo{year}{2018}).
\newblock \urlprefix\url{http://ol.osa.org/abstract.cfm?URI=ol-43-7-1487}.

\bibitem{huignard_phase_laser}
\bibinfo{author}{Brignon, A.} \& \bibinfo{author}{Huignard, J.~P.}
\newblock \emph{\bibinfo{title}{Phase Conjugate Laser Optics}}
  (\bibinfo{publisher}{Wiley}, \bibinfo{year}{2003}).

\end{thebibliography}

\end{document}


\maketitle


\section{Calculation of the Brillouin gain coefficient}

\begin{figure*}[b!]
        \centering
		\includegraphics[width = 11cm]{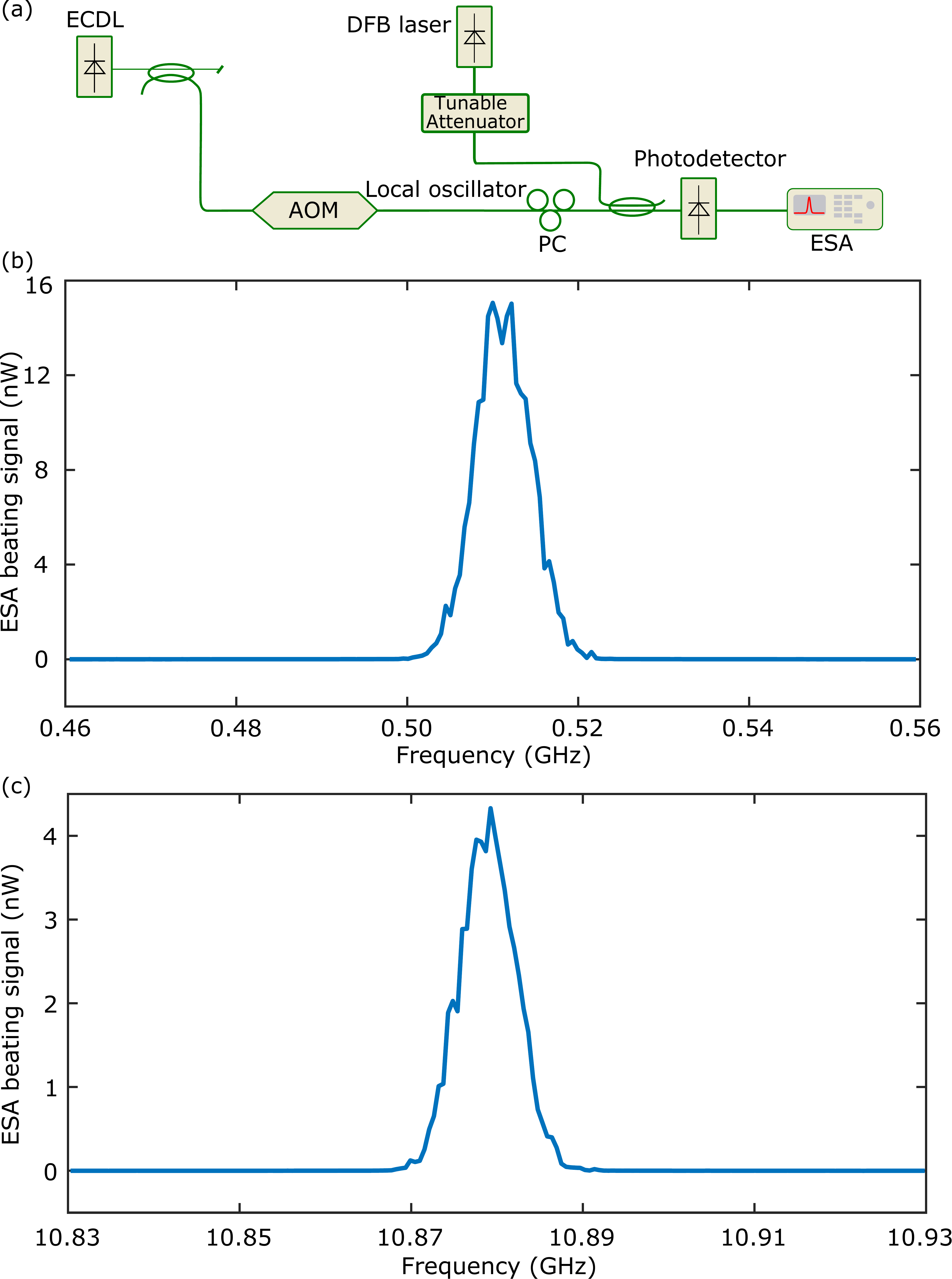}
		\caption{\textbf{Experimental beating signal between a distributed-feedback (DFB) laser and an external-cavity diode laser (ECDL).} (a) The beating experimental set-up. A DFB laser with power of -40 dBm (tuned by a tunable attenuator) beats with the ECDL which is used in the main manuscript. The beating signal is measured by a photodetector and an electrical spectrum analyser (ESA). The polarisation controller (PC) is used to maximise the beating signal. The resolution bandwidth and video bandwidth of the ESA are both 1 MHz. (b) Beating electrical power signal at low frequency range ($\sim$ 500 MHz) and (c) high frequency range ($\sim$ 10.9 GHz).}
		\label{fig_DFB_calibration}
\end{figure*}

The Brillouin gain coefficient and spectrum can be obtained by the spontaneous Brillouin scattering measurement \cite{boyd1990noise}. It has also been used for forward Brillouin scattering characterisation \cite{renninger_forward_2016,renninger_guided-wave_2016}. The Brillouin gain can be represented by $G = g_{\rm B} P_{\rm pump} L$, where $g_{\rm B}$ is the Brillouin gain coefficient in m$^{-1}$W$^{-1}$, $P_{\rm pump}$ is the pump optical power input to the fibre, and $L$ is the fibre length. When $G \ll 1$, which is the case when the reflected light originates entirely from the scattering of the laser field by spontaneously (i.e., thermally) generated phonons, the total generated Stokes and anti-Stokes optical power is given by \cite{boyd1990noise}:

\begin{equation}
    P_{\rm Stokes} \cong P_{\rm anti-Stokes} \equiv \frac{\pi}{2} g_{\rm B} P_{\rm pump} L  (\overline{n}+1) h \nu_{\rm pump} \delta_{\rm \nu}
\end{equation}

\noindent{where $P_{\rm Stokes}$ and $P_{\rm anti-Stokes}$ are the total Stokes and anti-Stokes optical power, $\overline{n} = (e^{\frac{h \nu_{\rm B}}{KT}} - 1)^{-1}$ is the mean number of phonons per mode of the acoustic field, h is the Planck constant, $\nu_{\rm B}$ is the Brillouin frequency shift (i.e. resonant acoustic frequency), $K$ is the Boltzmann constant, $T$ is the temperature in K, $\nu_{pump}$ is the pump frequency, $\delta_{\nu}$ is the full-width half-maximum Brillouin linewidth.}

The reflected spontaneous Stokes and anti-Stokes spectrum are measured by the set-up shown in Fig. 4(a) in the main manuscript. The Stokes and anti-Stokes power are calibrated by beating a known power distributed feedback (DFB) laser with an external cavity diode laser (ECDL) and the calibration set-up is shown in Fig. \ref{fig_DFB_calibration}(a). In Fig. 4(a) in the main manuscript, we measure not only the evanescent Brillouin scattering in the nanofibre gas cell (beating frequency of the Stokes signal at $\sim$ 500 MHz) but also the Brillouin scattering in standard single-mode fibre (beating frequency of the Stokes signal at $\sim$ 10.9 GHz). In order to measure the different response of the photo-detector (Newport model 1544) at 500 MHz and 10.9 GHz, we tune the frequency difference between the DFB laser and the ECDL at these two frequency regions. The beating spectra at these two frequency regions are shown in Figs. \ref{fig_DFB_calibration}(c) and (d) respectively with the attenuator tuned such that the DFB laser light reaches the coupler with a power of -40 dBm. The reflected total Stokes and anti-Stokes power due to gas evanescent Brillouin scattering in the 10 cm nanofibre gas cell are 4.5 nW with a pump power of 90 mW. The reflected total Stokes and anti-Stokes power due to Brillouin scattering in the 53.5 m standard single-mode fibre (SMF) are 2.9 nW with a pump power of 80 mW. Note the pump power difference is due to the insertion loss of the nanofibre. The measured linewidth for the gas evanescent Brillouin scattering and SMF Brillouin scattering are 17.1 MHz and 25.3 MHz respectively. With all these parameters, the peak Brillouin gain coefficient for the nanofibre gas cell filled with 40 bar CO$_2$ and for the SMF are calculated to be 8.2 m $^{-1}$W$^{-1}$ and 0.24 m$^{-1}$W$^{-1}$ respectively. The estimated SMF Brillouin gain coefficient 0.24 m$^{-1}$W$^{-1}$ is in excellent agreement with the value known for the SMF \cite{motil_invited_2016}.

\section{Analysis of the Brillouin signal from the nanofibre and tapered regions}

Our taper fabrication process is based on 3 steps to obtain an adiabatic transmission and a very low loss \cite{Hoffman_2014}.
The length of the adiabatic transition region is estimated by subtracting the total length of tapered fibre and the length of uniform section. In our case, the length of homogeneous section is 100 mm and the total length of tapered optical fibre is 179.4 mm. This value is precise because it is given by the translation stage. The length of both adiabatic transition regions is consequently 79.4 mm $\pm$ 5 mm.
The very good agreement between integrated measurement and calculation of Brillouin scattering all along the length of tapered optical fibre represented in Fig. 5(b) in the main manuscript reinforces the estimation of nanofibre length.  
The distributed Brillouin measurement in silica tapered optical fibre demonstrates the good agreement between estimation and measurements of the different lengths of tapered optical fibre \cite{Chow_2018}.

\section{Brillouin frequency shift as a function of CO$_2$ pressure in the nanofibre gas cell}

\begin{figure*}[htb]
        \centering
		\includegraphics[width = 10cm]{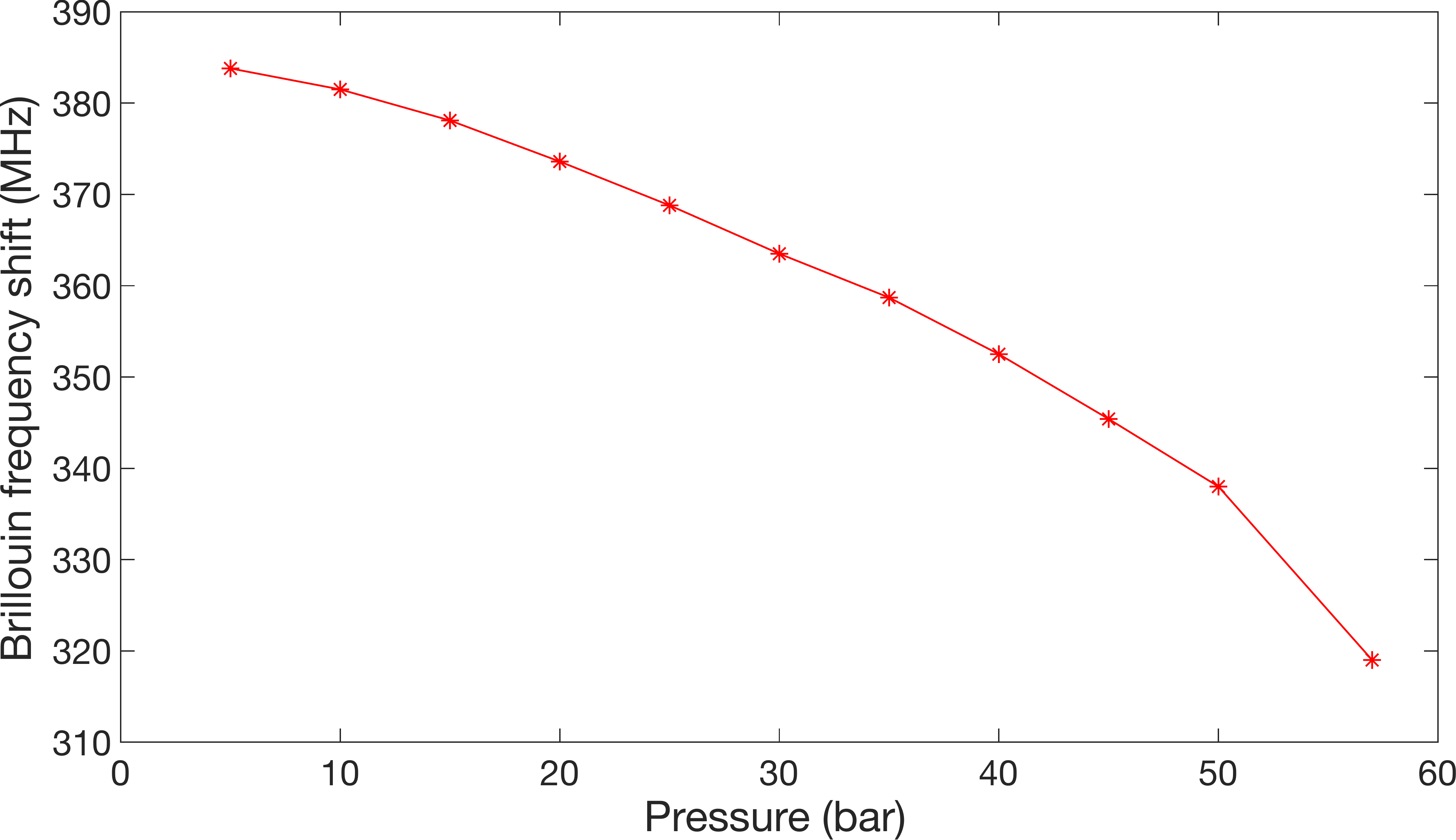}
		\caption{\textbf{Experimental Brillouin frequency shift for the nanofibre gas cell filled with different pressure of CO$_2$ gas.} }
		\label{fig_NF_BFS_Pressure}
\end{figure*}
Figure \ref{fig_NF_BFS_Pressure} shows the experimental Brillouin frequency shift for the nanofibre gas cell filled with different pressures of CO$_2$ gas. It demonstrates that our platform can be used for CO$_2$ gas pressure measurement by measuring the Brillouin frequency shift. Note that it can also be used for temperature measurement because for a fixed pressure, the acoustic velocity in the gas and hence the Brillouin frequency shift is related with the gas temperature with a coefficient of about 1.2 MHz/K.

\section{Theoretical analysis of the Brillouin scattering in a nanofibre liquid cell}

\begin{figure*}[hb!]
        \centering
		\includegraphics[width = 9.5cm]{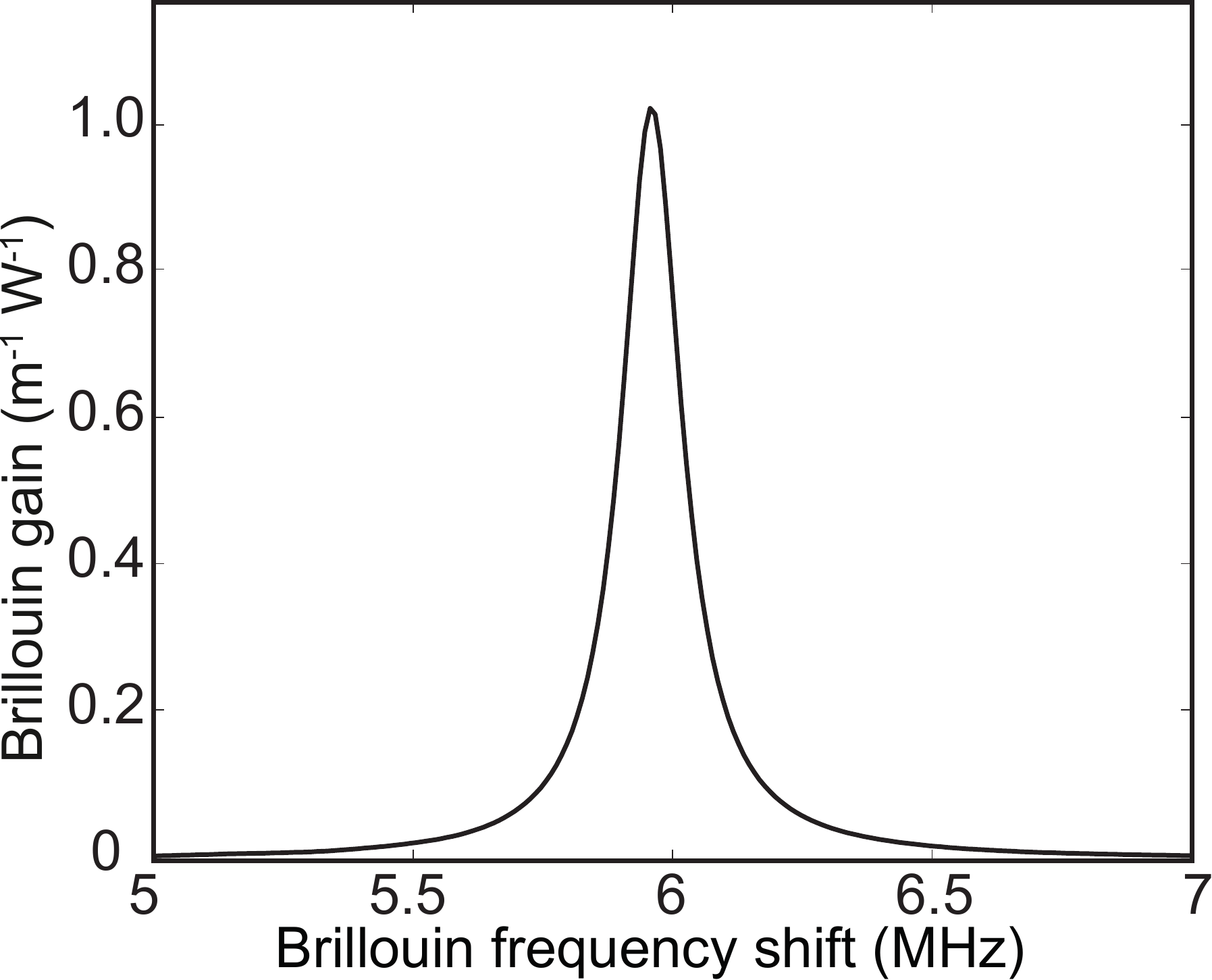}
		\caption{\textbf{Numerical calculation of evanescent Brillouin scattering in silica nanofiber surrounded by water.} }
		\label{fig_spectre_liq}
\end{figure*}

The Brillouin scattering generated by the evanescent field of a nanofibre in water is calculated and shown in Fig. \ref{fig_spectre_liq}. In the simulation, we use the parameters from \cite{huignard_phase_laser}. The optimal diameter of the nanofibre maximising the Brillouin scattering by the evanescent field in water is calculated to be 450 nm. The pump wavelength is 694 nm. The Brillouin frequency shift and Brillouin linewidth in water are 5.91 GHz and 371 MHz, respectively. In the calculation, the contribution from tapered regions is not included.

The peak Brillouin gain in Fig. \ref{fig_spectre_liq} is calculated to be 1 m$^{-1}$W$^{-1}$ which is equal to the peak Brillouin gain of a 740 nm nanofibre surrounded with 10 bar CO$_2$ and is 4 times higher than that of the standard single-mode fibre. This result shows good capability of our nanofibre waveguide for Brillouin spectroscopy and microscopy.

\bibliographystyle{naturemag}
\bibliography{supplement}